\documentclass[prd,nofootinbib]{revtex4}
\usepackage{epsfig,amsmath}

\def\Q1{{\bf Q}_1}

\def\d{{\rm d}}

\def\e{{\rm e}}

\def\gs{\gamma_S}
\def\tgs{$\gamma_S$ }
\def\gq{\gamma_q}
\def\ls{\lambda_S}
\def\vexp{$VT^3 \, \exp[-0.7 \, {\rm GeV}/T]$}

\def\ss{${\rm s}\bar{\rm s}\;$}
\def\ssb{\langle {\rm s}\bar{\rm s}\rangle}
\def\uub{\langle {\rm u}\bar{\rm u}\rangle}
\def\ddb{\langle {\rm d}\bar{\rm d}\rangle}

\def\kpi{$\langle {\rm K}^+ \rangle / \langle \pi^+ \rangle$}
\def\kmpim{$\langle {\rm K}^- \rangle / \langle \pi^- \rangle$}

\def\agev{$A$ GeV}
\def\agevt{$A$ GeV }

\def\snn{\sqrt{s}_{NN}}

\def\SHMgs{SHM($\gamma_S$)}

\begin{document}

\title{Energy and system size dependence of chemical freeze-out in relativistic
nuclear collisions.} 

\author{F. Becattini}\affiliation{Universit\`a di 
 Firenze and INFN Sezione di Firenze, Florence, Italy} 
\author{J. Manninen}\affiliation{University of Oulu, Oulu, Finland}
\author{M. Ga\'zdzicki}\affiliation{Institut f\"ur  Kernphysik, Universit\"at 
Frankfurt, Frankfurt, Germany and \'Swi\c{e}tokrzyska Academy, Kielce, Poland}

\begin{abstract}
We present a detailed study of chemical freeze-out in p-p, C-C, Si-Si and Pb-Pb 
collisions at beam momenta of 158$A$ GeV as well as Pb-Pb collisions at beam 
momenta of 20$A$, 30$A$, 40$A$ and 80$A$ GeV. By analyzing hadronic 
multiplicities within the statistical hadronization model, we have studied
the parameters of the source as a function of the number of the participating 
nucleons and the beam energy. We observe a nice smooth behaviour of temperature,
baryon chemical potential and strangeness under-saturation parameter as a 
function of energy and nucleus size. Interpolating formulas are provided which
allow to predict the chemical freeze-out parameters in central collisions
at centre-of-mass energies $\snn \gtrsim 4.5$ GeV and for any colliding ions. 
Specific discrepancies between data and model emerge in particle ratios in 
Pb-Pb collisions at SPS between 20$A$ and 40$A$ GeV of beam energy which cannot be 
accounted for in the considered model schemes.
\end{abstract}

\maketitle
\section{Introduction}

The main goal of the ultra-relativistic nucleus-nucleus (AA) collisions programme 
is to create in terrestrial laboratories a new state of matter, the Quark-Gluon 
Plasma (QGP). The existence of this phase, where quarks and gluons are deconfined, 
i.e. can freely move over several hadronic distances, is a definite prediction of  
Quantum Chromo-Dynamics (QCD). In the search for QGP, signals in nucleus-nucleus 
collisions at different nucleon-nucleon centre of mass ($\snn$) energies have been 
studied: from few GeV at AGS to several hundreds of GeV recently attained in Au-Au 
collisions at RHIC. 

One of the main results is the surprising success of the statistical-thermal models 
in reproducing essential features of particle production \cite{heinz}. This model 
succeeds also in describing particle multiplicities in many kinds of elementary 
collisions \cite{beca}, suggesting that statistical production is a general property 
of the hadronization process itself \cite{beca,vari}. Furthermore, 
the statistical hadronization model (SHM) supplemented with the hydrodynamical 
expansion of the matter, to a large extent, also reproduces transverse momentum 
spectra of different particle species \cite{marco}. 

Hence, the SHM model proves to be a useful tool for the analysis of soft 
hadron production and particularly to study strangeness production, whose enhancement 
has since long been proposed as a signature of QGP formation. Furthermore, anomalies 
in the energy dependence of strangeness production have been predicted \cite{predi}
as a signature of deconfinement and have been indeed observed experimentally 
\cite{MarekPhi}.

In a recent paper of ours~\cite{last}, we have studied these topics in detail 
comparing different versions of the statistical model. More recently, new experimental 
results~\cite{NA49CCSiSi} of hadronic multiplicities at top SPS beam energy - i.e. 
158 \agev, corresponding to $\snn = 17.2$ GeV - with light ion collisions (C-C and 
Si-Si), as well as in Pb-Pb collisions at beam energies of 20 and 30 \agev -
corresponding to $\snn = 6.2$ and 7.5 GeV respectively - became available. 
The Pb-Pb data has been recently analyzed \cite{rafhorn} within a version of the 
statistical model including a light-quark non-equilibrium parameter, $\gamma_q$. 
In the present work, we study the energy and system size dependence of 
chemical freeze-out by performing a systematic analysis of the available data
within a more essential framework of the SHM, including only the strangeness 
under-saturation parameter $\gamma_S$. With the updated data sample, we also test 
a two-component version of the SHM, where particle production stems from the 
superposition of fully chemically equilibrated fireballs and single nucleon-nucleon 
collisions; we also briefly address, once again, the issue whether the non-equilibrium 
factor $\gamma_q$ is allowed. 

The paper is organized as follows: a brief discussion on the SHM is given in 
Sect.~2. In Sect.~3 the experimental data selected for the analysis and the results 
of the analysis in our main and alternative schemes of the SHM are given. In 
Sect.~4 we present and discuss the energy and system size dependence of the chemical 
freeze-out stage. A general discussion of our results and the observed deviations
from the data is given in Sect.~5. Conclusions are drawn in Sect.~6.

\section{The statistical hadronization model in heavy ion collisions}

The statistical hadronization model has been described in detail elsewhere 
\cite{last}. Here we briefly summarize its founding concepts and discuss the issue 
of full-phase-space versus midrapidity analysis. 

The main idea of the statistical hadronization model is that hadrons are formed 
in statistical equilibrium within extended excited regions called {\em fireballs} 
or {\em clusters}. Several fireballs or clusters are produced in a single collision 
as a result of a dynamical process, each having a definite total momentum, charges 
and volume. 

In principle, the overall particle multiplicity should be calculated by adding
those relevant to single clusters and folding with the probability distribution 
of volumes, masses (being Lorentz invariants, they are independent of clusters' 
momenta) and charges. Unfortunately this calculation is not possible within the 
SHM alone, because fluctuations of volumes, charges and masses of clusters, being
governed by the previous dynamical process, are not known. However, if they happen
to be the same as those relevant to the random partitioning of one large cluster - 
defined as {\em equivalent global cluster} (EGC) - having as volume the sum of all 
clusters' rest frame volumes - the overall particle multiplicities turn out to be 
simply those the EGC \cite{last}, which has charges equal to the sum of single 
clusters' charges in compliance of conservation laws. The straightforward consequence
of such an assumption is that, in order to calculate full phase space mean 
multiplicities, one can use a simple single-cluster (i.e. the EGC) formula. Of 
course, this is just a phenomenological simplifying assumption and it should 
{\em not} be expected to be fully true; deviations from this simple picture 
should show up as second-order deviations between data and model.

It is very important to emphasize that the EGC picture takes advantage of the 
independence of particle multiplicity (as well as any other Lorentz scalar) on
clusters' momenta and, consequently, on any dynamical charge-momentum correlation.
This means that the hypothesis of single EGC calculation may hold even if clusters
with large baryon number (and density) are likely to have large rapidity.

To show this in simplest terms, let us consider events with $N$ clusters
and let $w(Q_1,y_1;\ldots;Q_N,y_N)$ the probability of having the $i^{\rm th}$ 
cluster with charge $Q_i$ at rapidity $y_i$. The overall average particle multiplicity 
of the species $j$ reads:
\begin{equation}\label{general}
 \langle\langle n_j \rangle\rangle = \sum_{Q_1,\ldots,Q_N} 
 \int \d y_1 \ldots \d y_N \; w(Q_1,y_1;\ldots;Q_N,y_N) 
 \langle n_j \rangle (Q_1,y_1;\ldots;Q_N,y_N),
\end{equation}
where $\langle n_j \rangle (Q_1,y_1;\ldots;Q_N,y_N)$ is the average multiplicity
of the species $j$ for a particular charge and rapidity configuration. Now, being 
a Lorentz scalar, $\langle n_j \rangle$ can only depends on charges, hence in fact 
$\langle n_j \rangle = \langle n_j \rangle (Q_1,\ldots,Q_N)$. Therefore:
\begin{equation}\label{special}
 \langle\langle n_j \rangle\rangle = \sum_{Q_1,\ldots,Q_N} 
 W(Q_1,\ldots,Q_N) \langle n_j \rangle (Q_1,\ldots,Q_N)
\end{equation} 
where: 
$$
W(Q_1,\ldots,Q_N)= \int \d y_1 \ldots \d y_N \; w(Q_1,y_1;\ldots;Q_N,y_N) .
$$
Thus, in order to reduce to the simple EGC case, we have to introduce some hypothesis
on the form of the {\em marginal} distribution of charges $W(Q_1,\ldots,Q_N)$ 
in Eq.~(\ref{special}) and we do not need to deal anymore with the full distribution 
of charges and rapidities $w$ in Eq.~(\ref{general}); charge-rapidity correlations 
have disappeared. 
 
In a different approach, used in several works, the formulas of the statistical 
model are assumed to apply only to the {\em average} (with respect to all kinds of 
fluctuations) fireball at midrapidity. The charge probability distribution 
$\omega(Q,y)$ of a fireball at rapidity $y$ is obtained from the general $w$ in 
Eq.~(\ref{general}) by integrating over all rapidities:
\begin{equation}
 \omega(Q,y) \propto \sum_{i=1}^N \sum_{Q_1,\ldots,Q_N} \delta_{Q,Q_i}
 \int \d y_1 \ldots \d y_N \; w(Q_1,y_1;\ldots;Q_N,y_N) \delta (y-y_i)
\end{equation}
Thus, in order to perform a statistical model fit, one would need to assume only 
that charge fluctuations at midrapidity, i.e. $\omega(Q,0)$, are not too large, which 
is apparently a less restrictive requirement than that needed to ensure the validity 
of the EGC approach. In order to estimate the parameters of the average central 
fireball, the idea is then to use particle yields and ratios measured at midrapidity 
rather than in full phase space. However, these yields do not single out the production 
from the central fireball as also nearby clusters contribute
to them and, if they have quite different mean charges, one would not get the 
desired result. Moreover, if a single cluster was formed at rest in the centre-of-mass 
frame, a cut on a midrapidity window ($\Delta y \approx 1$) would introduce a bias in 
the estimation of thermal parameters because rapidity distributions are narrower 
for more massive particles (see e.g. discussion in ref.~\cite{last}). The general 
effect would be an increase in the fitted temperature and strangeness under-saturation 
parameter $\gs$, what is indeed the case in an analysis at SPS using midrapidity 
densities \cite{pbmsps}. In fact, narrower rapidity distributions for more massive 
particles have been measured up to top SPS energy ($\snn \sim 20$ GeV).
For this approach to yield consistent results, one would need a distribution
of clusters with an approximately uniform charge and energy densities over a 
sufficiently large rapidity region, i.e. larger than the typical width of single 
cluster's particle rapidity spectrum. To show this, we start by writing the
rapidity spectrum of a particle as:
\begin{equation}
 \frac{\d N}{\d y} = \int_{-\infty}^{+\infty} \d Y \; \rho(Y) \,
 \frac{\d n}{\d y}\left( \mu(Y),T(Y),y-Y \right) \,
\end{equation}
where $\rho(Y)$ is the density of clusters at rapidity $Y$ and $\d n/ \d y$ is the 
primordial rapidity spectrum of particles emitted from the cluster with rapidity $Y$.
Changing the integration variable to $y' = y-Y$, one obtains:
\begin{equation}
 \frac{\d N}{\d y} = \int_{-\infty}^{+\infty} \d y' \; \rho(y-y') \,
 \frac{\d n}{\d y} \left( \mu(y-y'),T(y-y'),y' \right) .
\end{equation}
If $\rho(y-y')$ is approximately constant over a rapidity range (around $y$)
sufficiently larger than the width of $\d n/ \d y$, and so are $T(y-y')$ and 
$\mu(y-y')$, then:
\begin{equation}
\frac{\d N}{\d y} \approx \rho(y) \int_{-\infty}^{+\infty} \d y' \;
 \frac{\d n}{\d y}\left( \mu(y),T(y),y' \right) = \rho(y) \, n(T(y),\mu(y)) \,
\end{equation} 
hence the rapidity density turns out to be proportional to the number of
particles emitted from the average cluster at rapidity $y$. This condition
also implies that the measured $\d N/\d y$ is approximately constant over the same 
range.

However, this necessary condition is not met at centre-of-mass energies $\snn$ up 
to 20$A$ GeV, as measured rapidity distributions have a width not much larger than 
those from a single cluster at the kinetic freeze out. For instance, for pions at 
top SPS energy, the expected dispersion of the central cluster's rapidity 
distribution at 
a kinetic freeze-out temperature of $T \approx 125$ MeV, is about 0.8 while the 
dispersion of the actual distribution is $\simeq 1.3$ \cite{blume}. On the other 
hand, the measured width at $\snn = 200$ at RHIC is $\approx 2$ for pions \cite{brahms1} 
and antiparticle/particle ratio is apparently stable over a rapidity window of 
about 2 units around midrapidity \cite{brahms2}. These two observations indicate 
that extracting the characteristics of the average source at midrapidity by using 
midrapidity ratios is indeed possible at centre-of-mass energies ${\cal O}(100)$ 
GeV. Then, we conclude that the use of full phase space multiplicities is better 
suited over the energy range of AGS and SPS, that we are examining in the present
work, whilst midrapidity particle ratios can be used at RHIC energies to determine
the characteristics of the central source. Even though a statistical model fit to 
full phase space multiplicities at RHIC in principle may be attempted, provided
that the EGC condition applies, the presently available data set is not sufficient
to make a reliable assessment.

\section{The Data Analysis}

The experimental data set consists of measurements performed by NA49 collaboration 
in p-p (all inelastic reactions), C-C (15.3\% most central), Si-Si (12.2\% most 
central) collisions at beam momentum of 158\agev~as well as Pb-Pb collisions 
at beam momenta of 20$A$, 30$A$, 40$A$, 80$A$ and 158\agevt (corresponding to $\snn =$ 6.2,
7.5, 8.7, 12.3 and 17.2 GeV, respectively). The Pb-Pb data set corresponds to
7\% most central events except at top energy, where only 5\% thereof was selected. 
We also refit measurements in Au-Au collisions at 11.6\agevt of beam momentum
(corresponding to $\snn = 4.7$ GeV) by using the new hadronic data set input
\cite{pdg}.

The analysis is performed by searching the minima of the $\chi^2$:
\begin{equation}
 \chi^2 = \sum_i \frac{(N_i^{\rm exp} - N_i^{\rm theo})^2}{\sigma_i^2}
\end{equation}
in which $N_i$ is the full phase space of the $i^{\rm th}$ hadron species and 
$\sigma_i = \sqrt{(\sigma_i^{syst})^2 + (\sigma_i^{stat})^2}$ is the sum in 
quadrature of statistical and systematic experimental error.

The theoretical multiplicities are calculated by adding the primary multiplicities
to the contribution from secondary decays. In order to make a proper comparison
with the data, the decay chain is stopped so as to match the experimental definition 
of measured particle. For SPS data, weakly decaying particles are considered as
stable, except $\Lambda$ and $\bar{\Lambda}$ at 20$A$ and 30\agev, whereas for AGS 
Au-Au collisions, weak decays of hyperons and K$^0_S$ are performed. The final 
experimental and theoretical multiplicities used in our analysis are shown in 
Tables~\ref{Smallmult}, \ref{203040mult} and \ref{80158RHICmult}.  

We have performed the analysis with two independent programs, hereafter referred 
to as analyses A and B, to cross check the results and to assess the stability of 
the fits. We find only small discrepancies between the different analyses, mainly 
owing to a different method of treating hidden strangeness, branching ratios 
and a slightly different particle input. The effect of uncertainties in masses, 
widths and branching ratios have been shown to be negligible \cite{last} and 
the difference in results between the two analyses arises solely from the different 
selection of the included hadronic states.
The observed differences in the fit parameters between A and B are of the order 
of the fit errors and they may be considered as an estimate of the systematic error 
due to uncertainties in the implementation of the model.

We give two different errors for the fitted parameters and derived quantities
in Table~\ref{fit}, the first being the error coming out from the fitting 
program (inferred from the analysis of the $\chi^2=\chi^2_{\rm min} +1$ level 
contours) while the second one is the error rescaled by a factor 
$\sqrt{\chi^2_{\rm min}/dof}$ where $dof$ is the number of degrees of 
freedom. We deem that the latter is a more realistic uncertainty on the parameters 
because of the ``imperfect'' $\chi^2_{\rm min}/dof$ values (for further details, 
see~\cite{pdg,last}), and thus the rescaled errors are used in all plots in this paper.

\begin{table}
\caption{Summary of fitted parameters in nuclear collisions at AGS and SPS energies
in the framework of the SHM($\gs$) model. Also quoted strangeness 
chemical potential, minimum $\chi^2$'s, the estimated radius of the EGC and the $\ls$ 
parameter (see Sect.~3). The re-scaled errors, (see text) are quoted within brackets. 
For p-p at 158$A$ GeV of beam energy, we have fitted mean number of $s\bar{s}$ pairs 
(analysis A), and fitted \tgs (analysis B).}
\label{fit}
\vspace{0.5cm}
\begin{tabular}{|c|c|c|c|c|}
\hline
 Parameters     & Main analysis A & Main analysis B & Main analysis A & Main analysis B\\
\hline
%
%
\multicolumn{1}{|c|}{} & \multicolumn{2}{|c|}{p-p 158\agevt (C ensemble)} & 
\multicolumn{2}{|c|}{Au-Au 11.6\agevt (GC ensemble)} \\
\hline
$T$ [MeV]       & 181.5$\pm$3.4$^{a}$    & 178.2$\pm$4.8 (5.9)     & 118.7$\pm$2.7 (3.1)    & 119.2$\pm$3.9 (5.3) \\
$\mu_B$ [MeV]   &                        &                         & 554.4$\pm$11.3 (13.0)  & 578.8$\pm$15.4 (20.9)  \\
$\gs$           & 0.461$\pm$0.020$^{a,b}$& 0.446$\pm$0.018 (0.023) & 0.640$\pm$0.060 (0.069)& 0.768$\pm$0.086 (0.116) \\
\vexp           & 6.2$\pm$0.5$^{a,c}$    &0.127$\pm$0.005  (0.006) & 1.99$\pm$0.17 (0.20)   & 1.47 $\pm$0.18 (0.25) \\
\hline
$\chi^2$/dof    & 8.4/10$^{a}$           &    10.8/7               & 4.0/3                  & 5.5/3 \\
$R$ [fm]        &                        & 1.28$\pm$0.08 (0.10)    & 9.25$\pm$0.60 (0.69)   & 8.28$\pm$0.71 (0.96) \\
$\lambda_S$     & 0.266$\pm$0.019        & 0.195$\pm$0.005 (0.006) & 0.380$\pm$0.050 (0.058)& 0.489$\pm$0.083 (0.11) \\
\hline
%
%
\multicolumn{1}{|c|}{} & \multicolumn{2}{|c|}{C-C 158\agevt  (S-canonical ensemble)} & 
\multicolumn{2}{|c|}{Si-Si 158\agevt (S-canonical ensemble)} \\
\hline
$T$ [MeV]     & 166.0$\pm$4.4 (4.4)     & 166.1$\pm$4.2  & 162.2$\pm$4.9 (7.9)     & 163.3$\pm$3.0 (4.1) \\
$\mu_B$ [MeV] & 262.6$\pm$12.8 (12.9)   & 249.0$\pm$12.6 & 260.0$\pm$11.1 (17.9)   & 246.4$\pm$11.0 (15.1)\\
$\gs$         & 0.547$\pm$0.041 (0.041) & 0.578$\pm$0.043& 0.621$\pm$0.047 (0.076) & 0.668$\pm$0.049 (0.067)\\
\vexp         & 0.89$\pm$0.06  (0.06)   & 0.83$\pm$0.05  & 2.22$\pm$0.14 (0.22)    & 2.07$\pm$0.13 (0.18) \\
\hline
$\chi^2$/dof  & 4.1/4                   & 3.4/4          &  10.4/4                 & 7.6/4         \\
$R$ [fm]      & 2.89$\pm$0.19 (0.19)    & 2.82$\pm$0.19  & 4.15$\pm$0.30 (0.48)    & 3.99$\pm$0.19 (0.27) \\
$\lambda_S$   & 0.373$\pm$0.031 (0.032) & 0.364$\pm$0.034& 0.414$\pm$0.033 (0.054) & 0.418$\pm$0.036 (0.049)\\
\hline
%
%
\multicolumn{1}{|c|}{} & \multicolumn{2}{|c|}{Pb-Pb 20\agevt  (GC ensemble)} & 
\multicolumn{2}{|c|}{Pb-Pb 30\agevt (GC ensemble)} \\

\hline
$T$ [MeV]       & 131.3$\pm$2.3 (4.5)     & 135.8$\pm$3.2 (5.2)    & 140.1$\pm$1.6 (3.3)     & 144.3$\pm$1.9 (4.7)    \\
$\mu_B$ [MeV]   & 466.7$\pm$6.5 (12.9)    & 472.5$\pm$8.6 (13.7)   & 413.7$\pm$8.0 (16.3)    & 406.0$\pm$8.0 (19.1)   \\
$\gs$           & 0.773$\pm$0.037 (0.072) & 0.885$\pm$0.053 (0.086)& 0.773$\pm$0.041 (0.084) & 0.798$\pm$0.040 (0.099)\\
\vexp           & 4.41$\pm$0.23 (0.45)    & 3.88$\pm$0.26 (0.42)   & 6.91$\pm$0.40 (0.80)    & 6.52$\pm$0.35 (0.84)   \\
\hline
$\mu_S$ [MeV]   & 101.2                   & 114.2                  &  93.2                   & 99.8                   \\
$\chi^2$/dof    & 15.5/4                  & 10.3/4                 &  16.5/4                 & 23.0/4                 \\
$R$ [fm]        & 9.05$\pm$0.41 (0.80)    & 7.89$\pm$0.46 (0.73)   &  8.80$\pm$0.32 (0.64)   & 7.99$\pm$0.33 (0.79)   \\
$\lambda_S$     & 0.477$\pm$0.035 (0.069) & 0.586$\pm$0.056 (0.089)& 0.500$\pm$0.037 (0.073) & 0.517$\pm$0.039 (0.093)\\
\hline
%
%
\multicolumn{1}{|c|}{} & \multicolumn{2}{|c|}{Pb-Pb 40\agevt  (GC ensemble)} & 
\multicolumn{2}{|c|}{Pb-Pb 80\agevt (GC ensemble)} \\
\hline
$T$ [MeV]       & 146.1$\pm$2.2 (3.0)     & 143.0$\pm$2.3 (3.1)     & 153.5$\pm$2.5 (4.1)     & 149.9$\pm$3.2 (5.1)\\
$\mu_B$ [MeV]   & 382.4$\pm$6.8 (9.1)     & 380.8$\pm$6.6 (8.9)     & 298.2$\pm$5.9 (9.6)     & 293.8$\pm$6.9 (11.0)\\
$\gs$           & 0.779$\pm$0.033 (0.045) & 0.857$\pm$0.037 (0.050) & 0.740$\pm$0.024 (0.040) & 0.797$\pm$0.031 (0.049)\\
\vexp           & 8.75$\pm$0.40 (0.54)    & 7.57$\pm$0.35 (0.48)    & 15.25$\pm$0.61 (0.99)   & 13.53$\pm$0.64 (1.03)\\
\hline
$\mu_S$ [MeV]   & 89.5                    & 89.5                    & 69.6                    & 68.4          \\
$\chi^2$/dof    & 10.9/6                  & 11.0/6                  & 10.6/4                  & 10.2/4        \\
$R$ [fm]        & 8.53$\pm$0.35 (0.47)    & 8.59$\pm$0.35 (0.48)    & 9.05$\pm$0.38 (0.62)    & 9.23$\pm$0.44 (0.70) \\
$\lambda_S$     & 0.523$\pm$0.032 (0.043) & 0.513$\pm$0.031 (0.042) & 0.474$\pm$0.023 (0.038) & 0.443$\pm$0.021 (0.034) \\
\hline
%
%
\multicolumn{1}{|c|}{} & \multicolumn{2}{|c|}{Pb-Pb 158\agevt  (GC ensemble)} & 
\multicolumn{2}{|c|}{} \\
\hline
$T$ [MeV]       & 157.5$\pm$1.6 (2.5)    & 154.6$\pm$1.5 (2.7)     &   &    \\
$\mu_B$ [MeV]   & 248.9$\pm$5.7 (9.0)    & 245.9$\pm$5.6 (10.0)    &   &    \\
$\gs$           & 0.842$\pm$0.027 (0.042)& 0.941$\pm$0.030 (0.054) &   &    \\
\vexp           & 20.91$\pm$0.87 (1.39)  &  18.21$\pm$0.75 (1.35)  &   &    \\
\hline
$\mu_S$ [MeV]   & 59.3                   & 59.5                    &   &    \\
$\chi^2$/dof    & 22.5/9                 & 29.1/9                  &   &    \\
$R$ [fm]        & 9.42$\pm$0.27 (0.44)   & 9.42$\pm$0.27 (0.48)    &   &    \\
$\lambda_S$     & 0.526$\pm$0.020 (0.032)& 0.508$\pm$0.020 (0.036) &   &    \\
\hline 
\multicolumn{5}{|l|}{$a$ - In the fit A, the $\phi$ meson has been excluded from the data sample 
because it biased the fit towards}\\
\multicolumn{5}{|l|}{an exceedingly high temperature. The final $\chi^2$ does not then take into 
account the large deviation of $\phi$}\\
\multicolumn{5}{|l|}{meson yield.}\\
\multicolumn{5}{|l|}{$b$ - In the fit A, the $\gs$ parameter is to be replaced by the mean number 
                         $\ssb$ of strange quark pairs.}\\
\multicolumn{5}{|l|}{$c$ - In the fit A, the parameter \vexp~is to be replaced by $VT^3$.}\\
\hline   
\end{tabular}
\end{table}

\subsection{Main version}

\begin{center}
\begin{figure}[!ht]
$\begin{array}{c@{\hspace{1in}}c}
\epsfxsize=4.5in
\epsffile{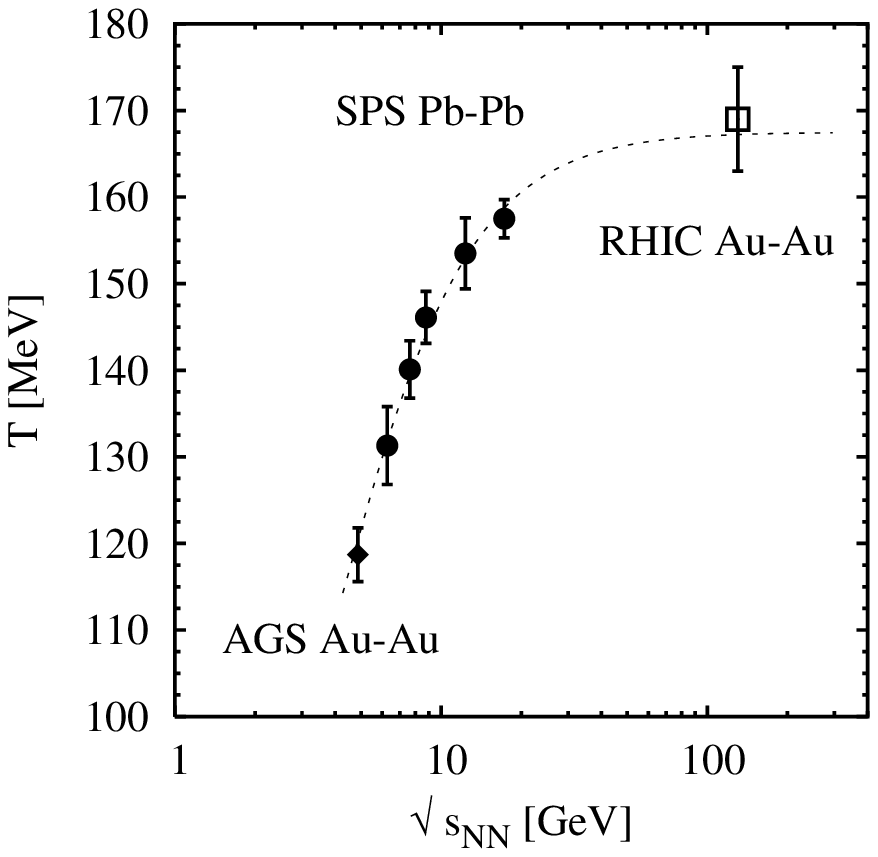} & \hspace*{-5cm}
        \epsfxsize=4.5in
        \epsffile{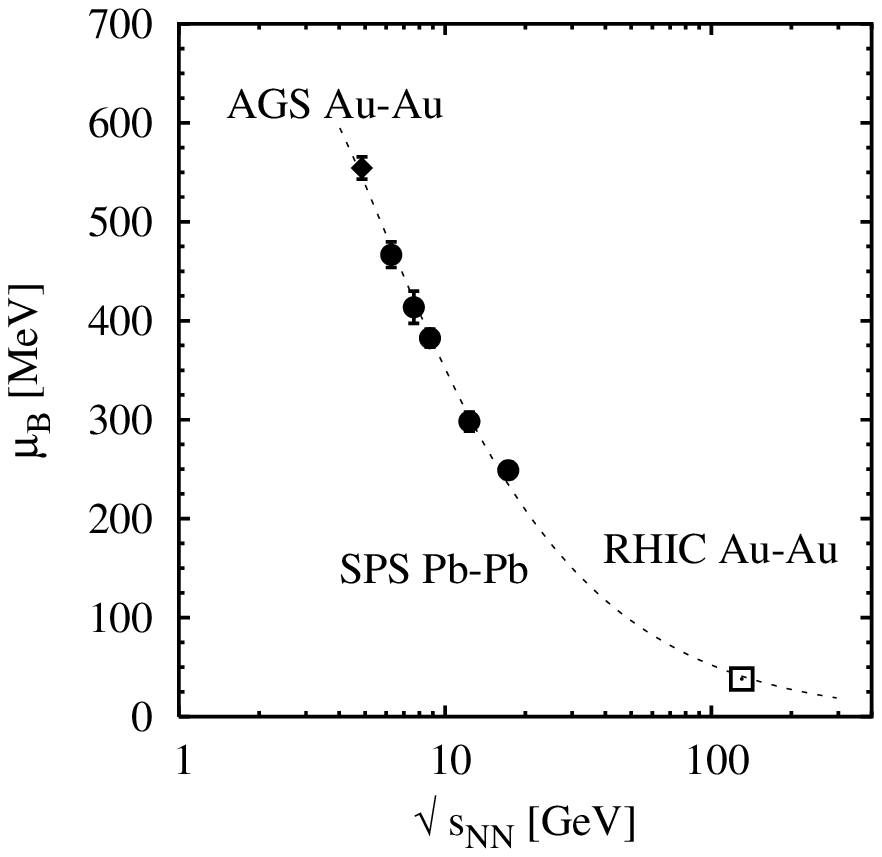} \\ 
\end{array}$
\caption{LEFT:
Fitted temperature at chemical freeze-out (analysis A) as a function of the N-N 
centre-of-mass energy in central Pb-Pb and Au-Au collisions. The dashed line is evaluated 
with the Eqs.~(\ref{tvsmu}) and (\ref{muvse}). The RHIC point, obtained with 
a fit to hadron ratios at midrapidity \cite{cleyRHIC} is shown as an open  square.\\ 
RIGHT:
Fitted baryon chemical potential at chemical freeze-out (analysis A) as a function 
of the N-N centre-of-mass energy in central heavy ion collisions. The dashed line 
is evaluated with the Eq.~(\ref{muvse}). The RHIC point, obtained with a fit to 
hadron ratios at midrapidity \cite{cleyRHIC} is shown as an open square.
\label{tmuen}}
\end{figure}
\end{center}

In our main version of the model we fit the parameters $T$, $V$, $\mu_B$ and the
strangeness under-saturation parameter $\gamma_S$; for the relevant formulas, see
ref.~\cite{last}. This version of the model is called \SHMgs.
The resulting \SHMgs~-parameters (see Table~\ref{fit} and Figs.~\ref{tmuen},~\ref{gslsen},
\ref{tmunp} and \ref{gslsnp}) are smoothly varying functions of the beam energy and no 
signs of anomalies are present. The fits have been performed in the supposedly 
sufficient statistical ensemble, i.e canonical (exact conservation of strangeness,
electric charge and baryon number) for p-p collisions, S-canonical (exact conservation 
of strangeness, but with grand-canonical treatment of electric charge and baryon 
number) in C-C and Si-Si collisions and grand-canonical (GC) for Pb-Pb collisions. 

Indeed, for C-C collisions, where the number of participants is quite small, we 
have cross-checked our main S-canonical calculation with a full canonical calculation
in which the baryon number has been set to the nearest integer number to the measured
average number of participants. Very little deviations have been found between the
two schemes for all particles, thus confirming the fitness of S-canonical ensemble.
On the other hand, the grand-canonical ensemble is not well suited both for C-C
and Si-Si collisions. By using the best fit parameters in the two pictures and 
comparing the fitted yields we have found that, for these systems, GC ensemble
overestimates the yield of multiply strange hyperons like $\Omega^-$ by 32\% 
and 14\% respectively with respect to S-canonical ensemble. This clearly indicates that 
S-canonical scheme is necessary.

The quality of the fits can be regarded as satisfactory overall, although not 
really good from a statistical point of view (except in Au-Au and C-C collisions)
as $\chi^2/dof$'s are generally of the order of 2-3. The worst fits are in Pb-Pb
collisions at 20$A$ and 30\agev, with $\chi^2/dof \sim 3-4$; this owes to specific 
deviations involving strange particles, that will be discussed in detail later
in Sect.~5.  
 
A special mention is needed for p-p collisions. As has been mentioned, theoretical 
multiplicities have been calculated in the canonical ensemble, which is described 
in detail in ref.~\cite{becapt}. As far as strangeness under-saturation is concerned, 
in the analysis B, the usual parametrization with $\gs$ has been used. On the other
hand, in the analysis A, a parametrization described in ref.~\cite{becapt} has 
been used in which it is assumed that some number of \ss pairs, poissonianly 
distributed, hadronizes. 
The parameter  $\gs$ is thus replaced by the mean number $\ssb$ of 
these \ss pairs. In general, for p-p, it was not possible to achieve a satisfactory
fit in both models. We believe that this owes to the inadequacy of the canonical 
ensemble at this relatively low energy, where micro-canonical effects play a non-
negligible role \cite{becaferro}. For analysis A, a good fit has been obtained 
removing the $\phi$ meson from the data sample, whose predicted yield deviates 
from the data around 70\%; indeed, the inclusion of $\phi$ leads the fit towards 
exceedingly high temperatures. For analysis B, a fair fit has been obtained removing 
the $\Xi$'s and $\Omega$'s baryons from the data sample. Also in this case, the
discrepancy between central data value and model are considerable, but it should
be also noted that the likely micro-canonical effects are the largest for heavy baryons
in pp collisions \cite{becaferro}. In general, it seems that the parametrization
with $\ssb$ leads to a better agreement with the data in comparison with the
$\gs$ parametrization, according to a cross-test performed in the framework of
the analysis A.

\subsection{Proper volume}

We have amended our fits by implementing extended hadrons instead of point-like 
particles. Our calculations
follow the model in ref.~\cite{rischke}, with a primary average multiplicity of 
the species $j$ in the grand-canonical ensemble reading, in the limit of Boltzmann 
statistics:
\begin{equation}\label{eigen}
  \langle n_j \rangle = \frac{ \langle n_j \rangle_{\rm pl} \e^{-v_j \xi}}
  {1 + \sum_k (v_k/V) \langle n_k \rangle_{\rm pl} \e^{-v_k \xi}}
\end{equation}
where $\langle n_j \rangle_{\rm pl}$ is the hadron multiplicity in the point-like
case, $\xi$ is the solution of the equation:
\begin{equation}
  \xi = \sum_k \langle n_k \rangle_{\rm pl} \e^{-v_k \xi}
\end{equation}
and $v_j$ is the proper volume, or eigenvolume, of the hadron. This volume effectively 
introduces a repulsive hard-core interaction in the hadron gas equation of state, yet 
it is an unknown quantity and one has to make some assumptions to develop calculations. 
If it was the same for all hadrons, there would be no corrections to the intensive 
parameters, as the ratio between different species would be the same (this can be seen 
from Eq.~\ref{eigen}). We have tested the assumption of an eigenvolume $v_j$ proportional 
to the mass through a bag-like constant $B$. In this case, one expects an upward 
shift in the temperature $\Delta T \simeq T^2 B \xi$, since, for masses $m \gg T$ one 
has:
\begin{equation}
  \langle n_j \rangle \propto \left(\frac{m T}{2 \pi}\right)^{3/2}\e^{-m_j/T - B \xi m_j}
\end{equation}
with all other fit parameters almost unchanged. Indeed, we have found that, for several
reasonable values of the constant $B$, from 0.5 to 2 fm$^3$/GeV, the only effect
in the fit was a temperature raise by the expected amount without any decrease in 
the $\chi^2$, that is in the fit quality. We therefore conclude that the introduction 
of this effect, at least in the model of ref.~\cite{rischke}, does not entail an 
improvement of the agreement data-model with respect to the point-like picture. 

\begin{center}
\begin{table}[ht]
\caption{Fit results in central C-C, Si-Si and Pb-Pb collisions at 158$A$ GeV 
within the two-component model SHM(TC) and with $\langle N_c \rangle$ as free 
parameter (top section) as well as  with $\langle N_c \rangle$ fixed to 
the value calculated in the 
Glauber model (bottom section). The re-scaled errors (see text) are quoted within brackets.
\label{fittc}}
\vspace{0.5cm}
\begin{tabular}{|c|c|c|c|}
\hline
  Parameters          & C-C, canonical ensemble & Si-Si, S-canonical ensemble & Pb-Pb, GC ensemble \\
\hline 
$T$ [MeV]             & 172.4$\pm$11.8 (12.6)  &  162.0$\pm$7.6   &   153.9$\pm$1.5 (2.5)\\
$\mu_B$ [MeV]         &                        &  234.4$\pm$22.5  &   240.8$\pm$6.9 (11.8) \\  
$\gs$                 & 1.0 (fixed)            &  1.0 (fixed)     &   1.0 (fixed)  \\
\vexp                 & 0.23$\pm$0.034 (0.037) &  0.91$\pm$0.11   &   16.11$\pm$0.57 (0.97)\\
$\langle N_c \rangle$ & 6.0$\pm$0.4 (0.4)      &  11.4$\pm$1.8    &   25$\pm$10 (16) \\
\hline
$\chi^2$/dof          & 5.8/4                  &  1.0/4           & 26.3/9 \\
\hline    
  Parameters          & C-C, S-canonical ensemble & Si-Si, S-canonical ensemble & Pb-Pb, GC ensemble \\
\hline 
$T$ [MeV]             & 161.0$\pm$9.1 (25.9)   &  151.1$\pm$7.1 (16.4)   &   154.7$\pm$1.5 (2.9)\\
$\mu_B$ [MeV]         & 315$\pm$18 (50)        &  285$\pm$13 (31)        &   261.6$\pm$2.6 (4.6) \\  
$\gs$                 & 1.0 (fixed)            &  1.0 (fixed)            &   1.0 (fixed)  \\
\vexp                 & 0.31$\pm$0.028 (0.071) &  1.19$\pm$0.06 (0.12)   &   16.54$\pm$0.44 (0.84)\\
$\langle N_c \rangle$ & 2.67 (fixed)           &  5.49 (fixed)           &   17.6 (fixed) \\
\hline
$\chi^2$/dof          & 32/5                   &  21.7/5                 &   36.3/10 \\
\hline    
\end{tabular}
\end{table}
\end{center}

\subsection{Two-component model}
\label{SHM_TC}

In this picture, henceforth referred to as SHM(TC), the observed hadron production 
is assumed to stem from the superposition of two components (TC): one originated 
from one or more fireballs in full chemical equilibrium and another component 
from peripheral single nucleon-nucleon collisions where final particles escape 
interaction region. The idea of this model is to ascribe the observed 
under-saturation of strangeness in heavy ion collisions to the N-N component, leaving 
the large fireballs at complete equilibrium, i.e. with $\gs=1$. 
Of course, this is possible provided that the mean number of single nucleon-nucleon 
collisions ($\langle N_c \rangle$) is sufficiently large. We note that similar 
approaches have been proposed in literature where the role of the second component 
beside the main fireball is played by smaller peripheral fireball with different 
thermodynamical parameters or by a collection of clusters \cite{hohne}.
It should be mentioned that the SHM(TC) picture is quite a simplified one: it
is assumed that particles emerging from N-N collisions decouple without further
re-interaction, whereas collisions in the core of the system eventually lead to a 
hadron gas in perfect equilibrium. These are sharp approximations that would 
certainly need a refinement in more accurate studies. 

In the SHM(TC) model, the overall hadron multiplicity can be written then as:
\begin{equation}
 \langle n_j \rangle = \langle N_c \rangle \langle n_j \rangle_{NN} +
 \langle n_j \rangle_V
\end{equation}
where $\langle n_j \rangle_{NN}$ is the average multiplicity of the $j^{\rm th}$
hadron in a single N-N collision and $\langle n_j \rangle_V$ is the average 
multiplicity of hadrons emitted from the equilibrated fireball with $\gs=1$. The 
$\langle n_j \rangle_{NN}$ term can be written in turn as:
\begin{equation}
 \langle n_j \rangle_{NN} = \frac{Z^2}{A^2} \langle n_j \rangle_{pp} +
 \frac{(A-Z)^2}{A^2} \langle n_j \rangle_{nn} + \frac{2Z(A-Z)}{A^2}
 \langle n_j \rangle_{np} \
\end{equation}
To calculate $\langle n_j \rangle_{NN}$ we have used the statistical model and
fitted p-p full phase space multiplicities measured at the same beam energy by 
the same NA49 experiment (see Table~\ref{Smallmult}). For n-p and n-n collisions, the 
parameters of the statistical model determined in p-p are retained and the 
initial quantum numbers are changed accordingly. 

We have fitted $T$, $V$, $\mu_B$ of the equilibrated fireballs and 
$\langle N_c \rangle$ by using NA49 data in C-C, Si-Si and Pb-Pb collisions at 158\agev~
in the analysis A. It should be mentioned that, in this fit, the systematic error on 
the refitted parameters owing to the uncertainty on statistical model parameters in N-N 
collisions (i.e. the errors quoted in Table~\ref{fit} for p-p collisions), used as 
input, has been disregarded. The resulting fit parameters are shown in Table~\ref{fittc}.
For the Pb-Pb system, the fit quality, as well as the obtained values of $T$ and $\mu_B$, 
are comparable to the main fit within the SHM($\gs$) model. The predicted number of
``single" N-N collisions is about 25. Thus, on average, only 310 nucleons out of 360 
participants contribute to the formation of large equilibrated fireballs.
In case of Si-Si, the fit quality is significantly improved compared to the main 
version of the statistical model, and the resulting fit parameters are comparable 
to the ones in Table~\ref{fittc}. The number of independent N-N collisions is  
12$\pm$4, suggesting that more than half of the 41 participating nucleons 
are colliding only once. 

Being so small, C-C system needs special treatment when two component model is 
applied. Since there are only 16 participants in the system, one has to calculate 
both components in full canonical ensemble. This means that one has to take 
explicitly into account all different proton-neutron configurations in the central 
fireball and extract those nucleons participating to the fireball from the 
'nucleon pool' available for the single N-N collisions. It seems that even C-C can 
be described with the two component model if the baryon charge in the central 
fireball is of the order B=n+p=4. The resulting  $\chi^2/dof \simeq 5.8/4$ with 
all the different n+p=4 combinations in the completely equilibrated fireball, i.e. 
the fit quality is worse than with the main version of the statistical 
model, but the overall fit quality is acceptable anyhow.

To cross-check these results, we have calculated the number of single N-N collisions 
from the Glauber model and repeated the fits (in the analysis B) the same way by fixing 
$\langle N_c \rangle$ as that coming from the Glauber calculations. We first 
implemented a Monte-Carlo calculation of the Glauber 
model as follows: 
\begin{enumerate}
\item{} at a given impact parameter, the number of collisions of each projectile 
nucleon is randomly extracted from a Poisson distribution whose mean is the 
product of the thickness function times the inelastic nucleon-nucleon cross-section; 
\item{} then, the number of collisions undergone by each nucleon belonging to
the target nucleus is randomly extracted from a multinomial distribution 
constrained with the total number of collisions as determined in the previous step 
and whose weights are proportional to the product of their relevant thickness 
function times the inelastic nucleon-nucleon cross-section;
\item{} for each generated event, uniformly distributed in the transverse impact
parameter plane, we keep track of the number of nucleons undergoing $0, 1, 2, \ldots$
collisions $N_0, N_1, N_2, \ldots$ in both projectile and target nucleus. 
\end{enumerate}

The thickness function has been calculated on the basis of a 
Woods-Saxon distribution:
\begin{equation}\label{ws}
  \frac{dN}{dr} = \frac{n_0}{1+{\rm e}^{(r-R)/d}} 
\end{equation}
with parameters quoted in Ref.~\cite{misko}:
$$
  n_0 = 0.17 \; {\rm fm}^{-3}; \;R= 1.12 A^{1/3}-0.86 A^{-1/3} \; {\rm fm};\;
d = 0.54  \; {\rm fm}.
$$
For each event, this Monte-Carlo calculation provides the number of nucleons 
$N_{1(a)}$ and $N_{1(b)}$ colliding once, in the projectile $a$ and target 
nucleus $b$ respectively. In general, these two numbers differ because nucleons 
from $a$ may collide once with a nucleon from $b$, which in turn collides with 
two or more nucleons from $a$. Therefore, the minimum between $N_{1(a)}$ and 
$N_{1(b)}$ is the maximal number of single nucleon-nucleon collisions per 
event, i.e.
\begin{equation}
  N_c \ge \min[N_{1(a)},N_{1(b)}] \equiv N_m
\end{equation}
The equality sign applies only if the $N_m$ nucleons from one nucleus 
collided with $N_m$ nucleons of the other, all of them among those undergoing
one collision. We assume this is always the case and hence take $N_m$ as a fair 
estimate of the number of single collisions. The mean number of $N_c$ is then
averaged over the most central bin defined by the distribution of the number of 
projectile spectators according to the NA49 centrality selection method.

We determined the mean number of single N-N collisions within the most central
bin to be 2.67, 5.49 and 17.6 for C-C, Si-Si and Pb-Pb respectively at a beam
energy of 158\agev. By fixing $N_c$ to the above numbers, we then fitted $V$, 
$T$ and $\mu_B$ of the equilibrated fireballs and found the results shown in 
Table~\ref{fit}. The fit quality is size-ably worse than in the previous case. 
Since the used number of collisions is actually an upper limit of single N-N collisions, 
we conclude that the SHM(TC) is disfavored by the data, if the Glauber model is 
assumed to be correct. 

\section{Energy and system size dependence} 

We are now in a position to study the dependence of chemical freeze-out
on beam energy and system size in central ultra relativistic heavy ion collisions. 
According to what has been discussed at the end of Sect.~2, for RHIC energy,
we will include the parameters determined with fits to midrapidity particle 
multiplicity ratios.

\begin{center}
\begin{figure}[!ht]
$\begin{array}{c@{\hspace{1in}}c}
\epsfxsize=4.5in
\epsffile{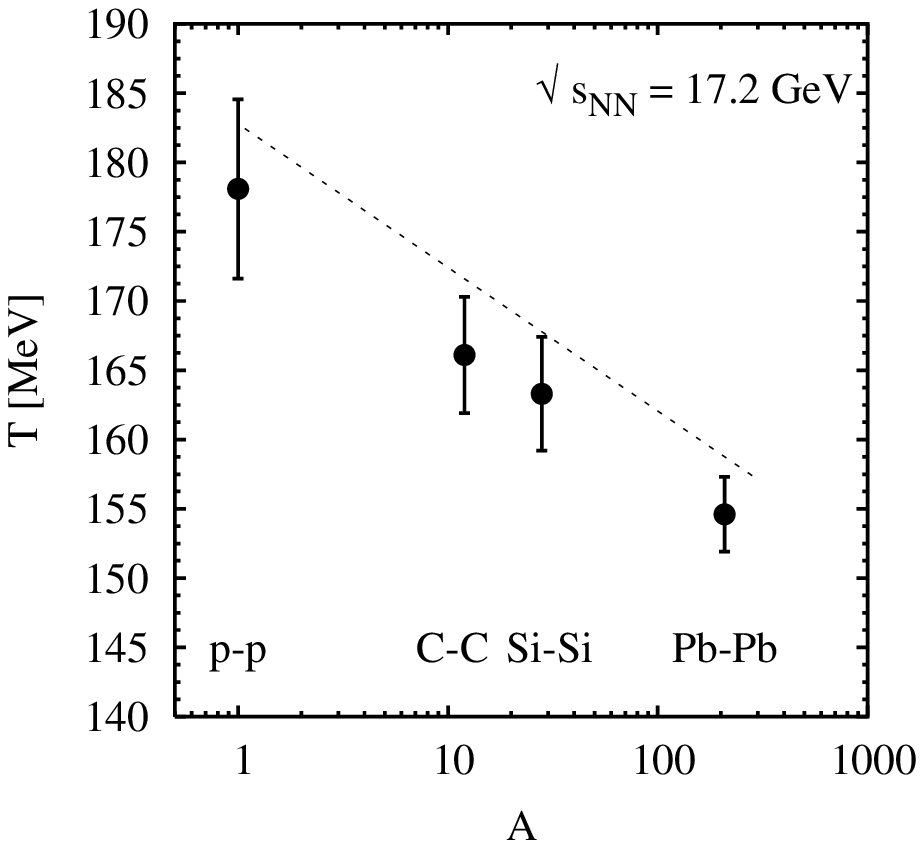} & \hspace*{-5cm}
        \epsfxsize=4.5in
        \epsffile{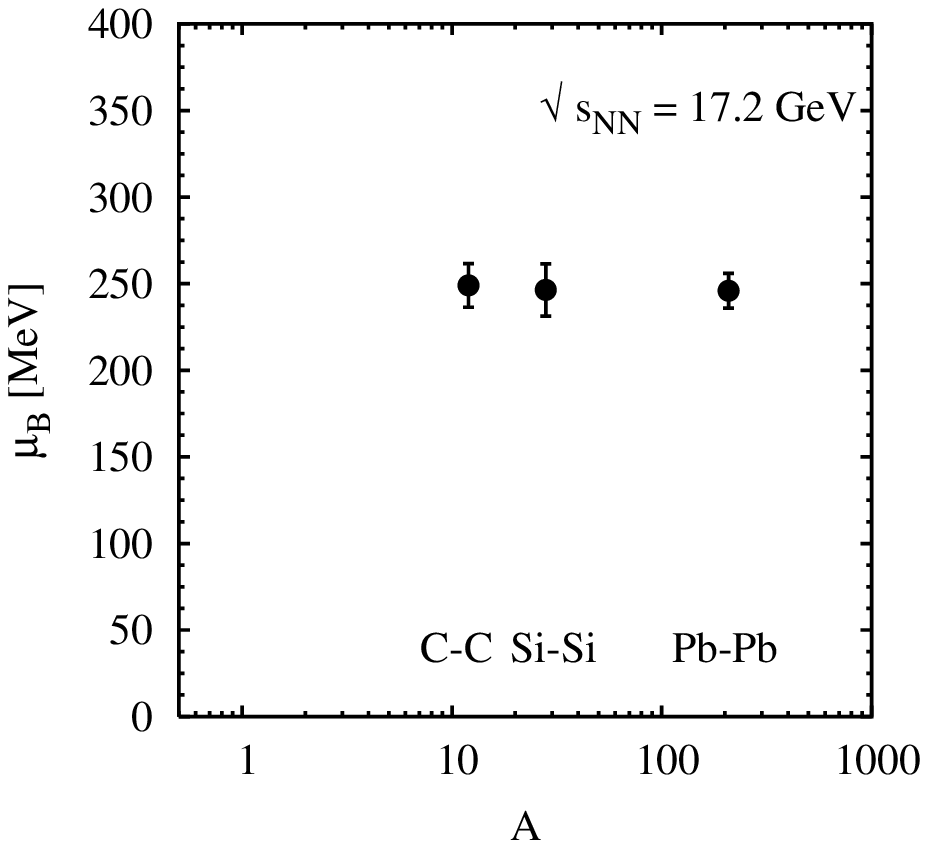} \\ 
\end{array}$
\caption{Fitted temperature (LEFT) and baryon chemical potential (RIGHT) at 
chemical freeze-out as a function of $A$ in central heavy ion collisions at
$\snn = 17.2$ GeV. From left to right: p-p, C-C, Si-Si and Pb-Pb points 
fitted in analysis B. The dashed line is evaluated with the Eq. (\ref{tvsall}) 
which has been fitted to the points of analysis A. The observed systematic
shift between data and interpolation is mainly a reflection of the slight 
difference ($\simeq 3-4$ MeV) between temperature in analyses A and B in 
Pb-Pb and p-p.  
\label{tmunp}}
\end{figure}
\end{center}

The first observation is that the chemical freeze-out of different colliding
systems at the same N-N centre-of-mass energy, i.e. C-C, Si-Si and Pb-Pb, seem 
to occur at similar values of the baryon chemical potential, namely 
$\mu_B\approx$~250 MeV (see Fig.~\ref{tmunp}). Such weak system size dependence 
of the baryon chemical potential has been already reported 
earlier~\cite{cley_systemsize}. On the other hand, systems with fewer participating 
nucleons seem to decouple at slightly higher temperature than heavy systems 
(see Fig.~\ref{tmunp}). Nevertheless, generally speaking, the freeze-out condition 
seems to be determined mostly by the beam energy and little by the number of 
participants, also in peripheral collisions~\cite{cley_systemsize,maiani}. 

\begin{center}
\begin{figure}[!ht]
$\begin{array}{c@{\hspace{1in}}c}
\epsfxsize=4.5in
\epsffile{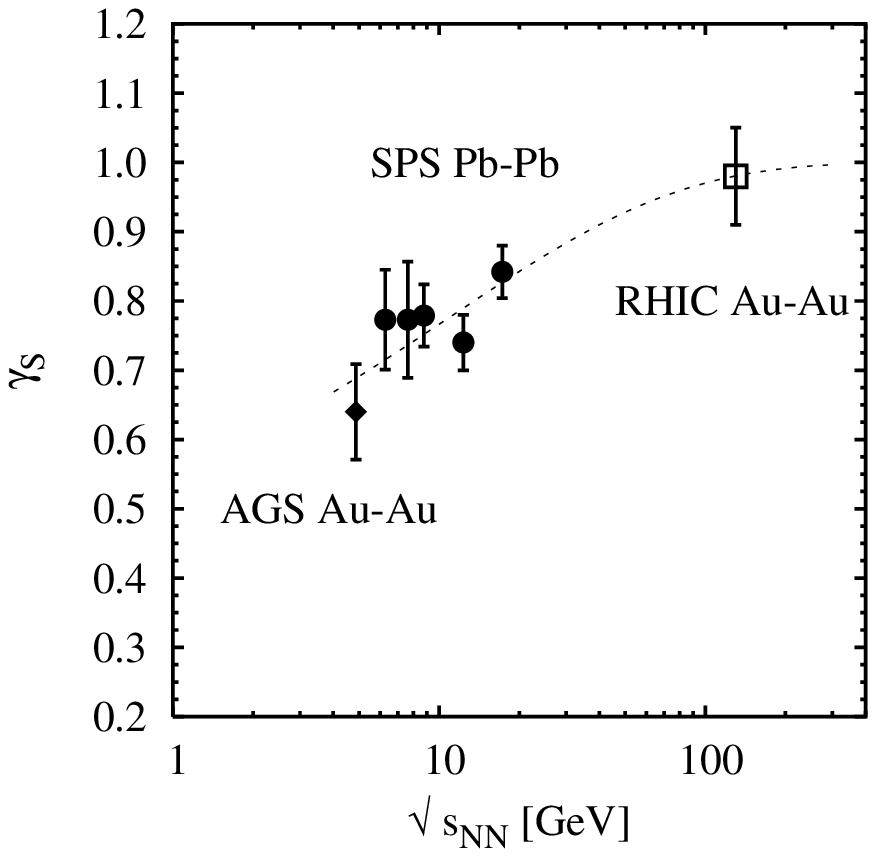} & \hspace*{-5cm}
        \epsfxsize=4.5in
        \epsffile{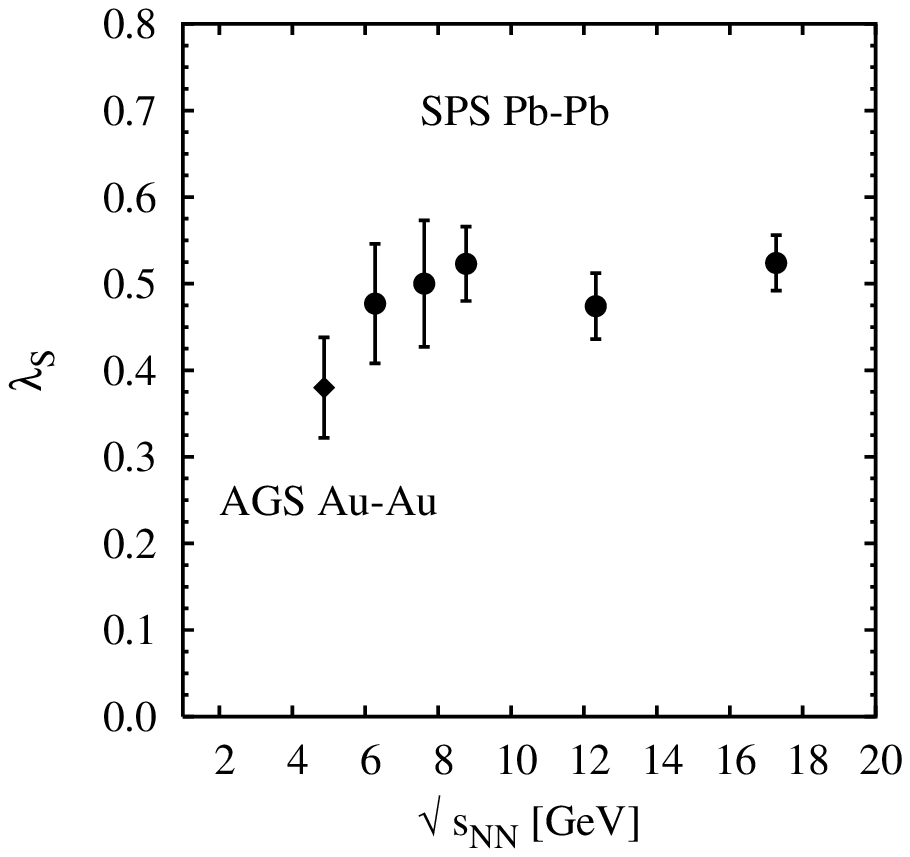} \\ 
\end{array}$
\vspace{-0.3cm}
\caption{LEFT:
Strangeness undersaturation parameter $\gamma_S$ at chemical freeze-out (analysis A) 
as a function of N-N centre-of-mass energy in central heavy ion collisions. 
The dashed line is evaluated with the Eq.~(\ref{gsvsall}). The RHIC point, obtained 
with a fit to hadron ratios at midrapidity \cite{cleyRHIC} is shown as a square dot.\\
RIGHT: The Wroblewski factor $\lambda_S$ (analysis A).
\label{gslsen}}
\end{figure}
\end{center}

The dependence of the freeze-out parameters on the N-N centre-of-mass energy
in heavier systems (Pb-Pb and Au-Au) is shown in Figs.~\ref{tmuen},\ref{gslsen}. 
The fitted points show a relatively strong dependence on energy, yet with a  smooth 
behaviour: lower energies always correspond to a lower central temperature value 
and a higher baryon chemical potential. 

\begin{center}
\begin{figure}[ht]
\epsfxsize=6.5in
\epsffile{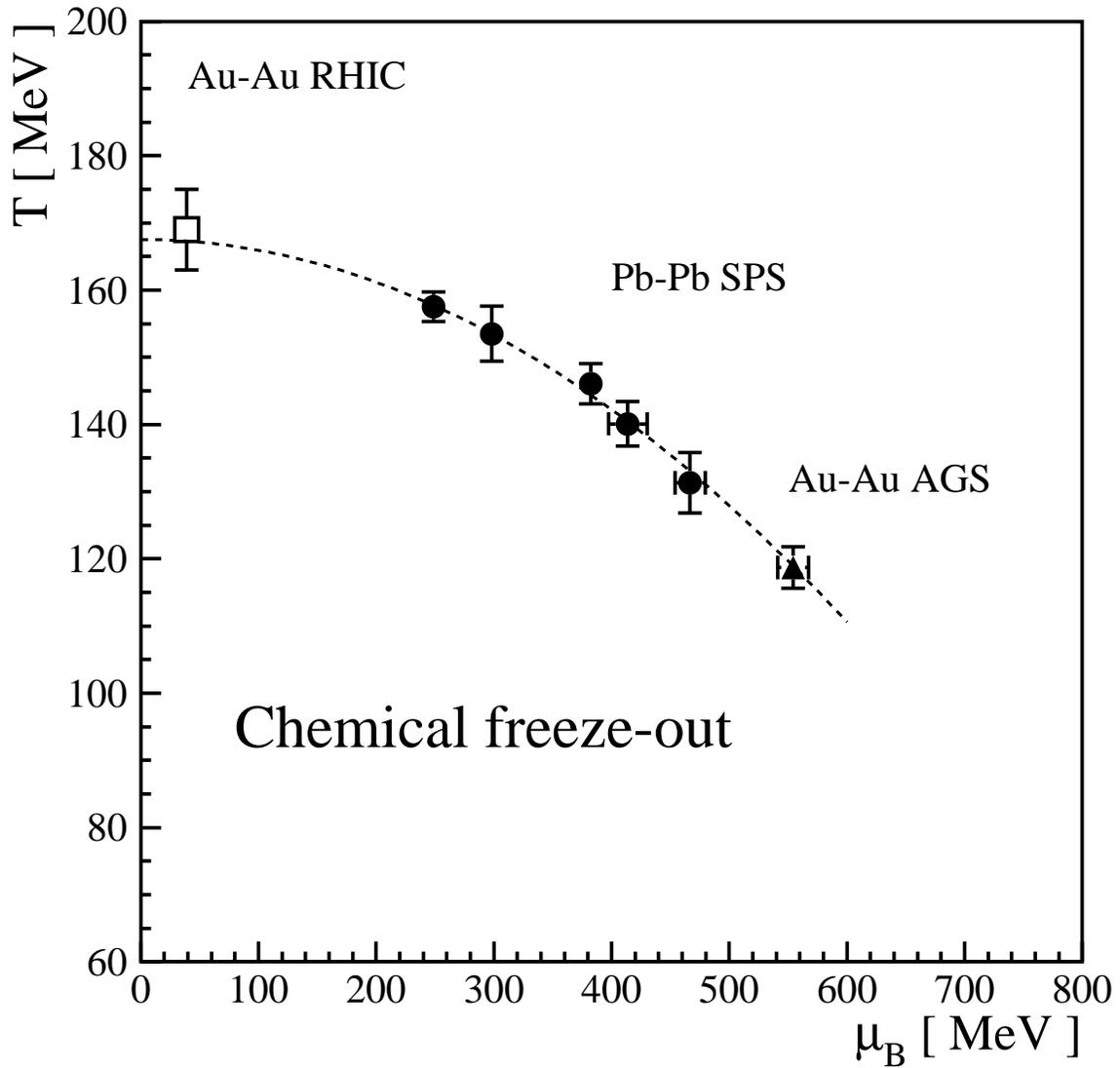} 
\caption{Chemical freeze-out points in the $\mu_B-T$ plane for 
central  Pb-Pb and Au-Au
collisions. The dashed line shows the parabolic interpolation Eq.~(\ref{tvsmu})
obtained by fitting SPS and AGS points. The RHIC $\snn=130$ GeV point, shown as an 
open square, has been taken from the recent analysis in ref.~\cite{cleyRHIC}.
\label{tmu}}
\end{figure}
\end{center}

A smooth dependence is also obtained for the chemical freeze-out in the 
$\mu_B-T$ plane, see Fig.~\ref{tmu}; all points lie, over the examined range of 
temperatures, on a parabola corresponding to chemical freeze-out 
condition $\langle E \rangle / \langle N \rangle\approx$ 1 GeV~\cite{cley1GeV}.
Owing to the slight difference in temperature, the lighter systems C-C and
Si-Si do not lie on the same curve and this leads to the conclusion that, at 
least in the model with $\gs$ only, freeze-out curve depends on colliding nucleus. 

\begin{center}
\begin{figure}[!ht]
$\begin{array}{c@{\hspace{1in}}c}
\epsfxsize=4.5in
\epsffile{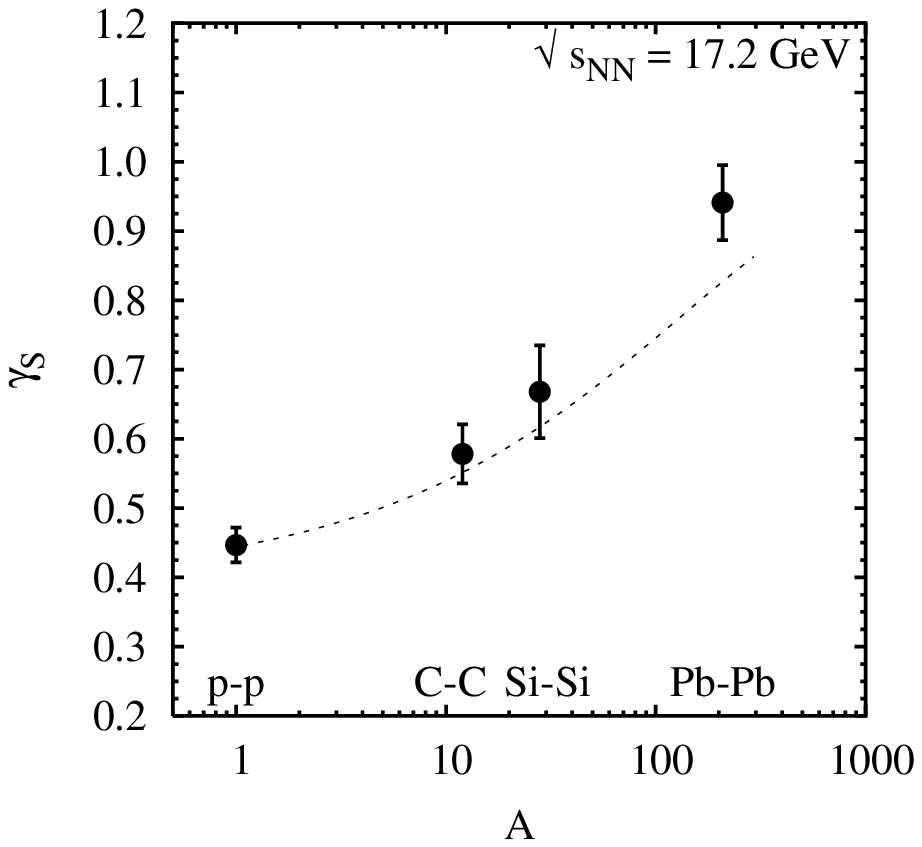} & \hspace*{-5cm}
        \epsfxsize=4.5in
        \epsffile{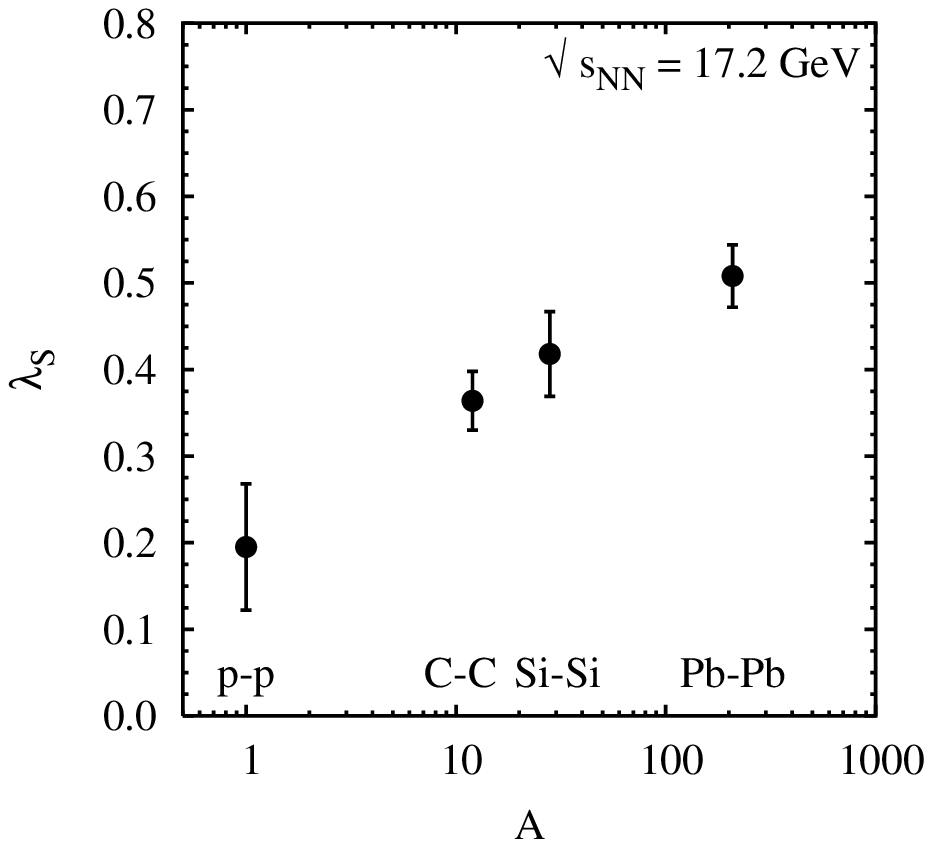} \\ 
\end{array}$
\vspace{-0.3cm}
\caption{Strangeness suppression factor $\gs$ (LEFT) and the Wroblewski factor 
$\lambda_S$ (RIGHT) at chemical freeze-out as a function of $A$ in central heavy 
ion collisions at $\snn=17.2$ GeV (analysis B). From left to right: p-p, C-C, Si-Si 
and Pb-Pb. The dashed line is evaluated with the Eq.~(\ref{gsvsall}). 
\label{gslsnp}}
\end{figure}
\end{center}

Conversely, the degree of chemical equilibration of strangeness seems to be 
strongly dependent on the number of participants, and much less on the energy. 
This has been pointed out in several studies of peripheral Pb-Pb collision 
systems~\cite{cley_systemsize,maiani}. In fact, from small to large systems, 
$\gamma_S$ increases monotonically from 0.45 in p-p to $\sim 0.8$ in Pb-Pb at 
the same beam energy, see Fig.~\ref{gslsnp}, while the dependence on energy is 
much milder, with a variation from 0.65 to 0.84 over the 4.7-17.2 GeV energy range. 
Therefore, strangeness is not in chemical equilibrium up to SPS and only 
for midrapidity yields at RHIC $\gs$ seemed to have attained one \cite{rhictm}.

Strong suppression in strangeness production in small systems can be seen also 
in the Wroblewski variable $\ls = 2 \ssb/(\uub+\ddb)$, the estimated 
ratio of newly produced strange quarks to u, d quarks at primary hadron level, 
shown in Figs.~\ref{gslsen},\ref{gslsnp}, and Table~\ref{fit}. The calculation of 
newly produced quark pairs is performed by using the statistical model best fit 
values of the various hadron multiplicities, so the obtained $\ls$ values are somehow 
model-dependent. 
Nevertheless, this variable shows a very similar behaviour as the \tgs parameter, 
indicating strong monotonic system size dependence in relative strangeness production.

\begin{center}
\begin{figure}[!ht]
$\begin{array}{c@{\hspace{1in}}c}
\epsfxsize=4.5in
\epsffile{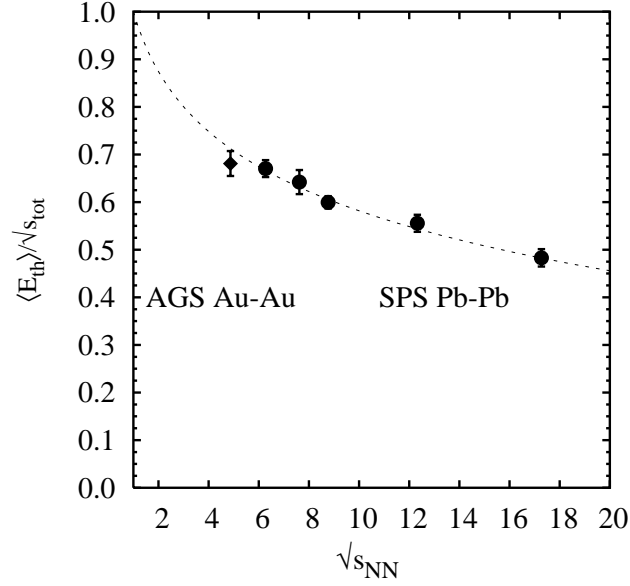} & \hspace*{-5cm}
        \epsfxsize=4.5in
\end{array}$
\vspace{-0.3cm}
\caption{Estimated fraction of the initial collision energy spent into
thermal energy content in central heavy ion collisions as a function of the N-N
centre-of-mass energy. The line is of the form $f(\snn) = 1-b\ln(\snn)$ and 
fitted to all points.
\label{eloss}}
\end{figure}
\end{center}

It is also interesting to estimate the thermal energy content of the hadronic matter 
at freeze-out. Indeed, the fraction of thermal energy to the total available energy 
decreases rapidly as a function of energy, as shown in Fig.~\ref{eloss}. 

\subsection{Interpolation of the statistical model parameters}

It is worth trying to summarize the amount of information we have collected on the 
statistical model parameters at freeze-out in the central collisions of different systems 
with some simple empirical interpolations. Some of them have already been proposed 
in previous works \cite{pbm,last}.

A satisfactory description of our Pb-Pb and Au-Au freeze-out points in the $T-\mu_B$ plane 
can be achieved by the simple parabolic fit:
\begin{equation}\label{tvsmu}
 T \approx T_0 - b \,\mu_B^2  \qquad T_0 = 167.5 \,{\rm MeV}  \;\;\; b = 0.1583 \,
 {\rm GeV}^{-2} .
\end{equation}
Note that the most recent determination~\cite{cleyRHIC} of $T$ and $\mu_B$ in central 
Au-Au collisions at $\snn =130$ GeV (169$\pm$6 MeV and 38$\pm$5 MeV, respectively)
also follows this dependence, see Fig.~\ref{tmu}. Similarly, for $\gs$, one can try to 
make an interpolation as a function of $\mu_B$ constrained by the requirement 
$\gs \rightarrow 1$ when $\mu_B \rightarrow 0$, i.e. full chemical equilibrium at very 
large energy. The assumed functional form is then:
\begin{equation}\label{gsvsmu}
 \gs = 1 - g \exp \left[ -\frac{a}{\mu_B/T} \right]
\end{equation} 
and the resulting parameters are $g=0.396$ and $a=1.23$. Both fits to Eqs.~(\ref{tvsmu})
and (\ref{gsvsmu}) have  $\chi^2/dof \approx 1$.

The next step is to provide an interpolation of $\mu_B$ as a function of $\snn$.
Once this was known, it would be possible to predict the value of $T$ and $\gs$ at
any collision energy through the Eqs.~(\ref{tvsmu}) and (\ref{gsvsmu}). The main advantage of
$\mu_B$ is that it is almost independent of the number of participants,
so that a single function $\mu_B=\mu_B(\snn)$ would apply to both light and heavy 
systems. We have interpolated $\mu_B$ with a function:
\begin{equation}\label{muvse}
 \mu_B = \alpha \frac{\log \snn}{(\snn)^\beta} \,
\end{equation} 
with $\alpha=2.06$ and $\beta=1.13$ and energy is given in GeV. This interpolation 
gives a very good description of the energy dependence of baryon chemical potential 
from $\snn=4.7$ to 130 GeV. 

We are now in a position to obtain the functions $T=T(\snn,A)$ and $\gs=\gs(\snn,A)$.
According to Eqs.~(\ref{tvsmu}) and (\ref{muvse}), we have:
\begin{equation}
  T = T_0 - C \left( \frac{\log \snn}{(\snn)^\beta} \right)^2 \,
\end{equation}
where $C = b \alpha^2 \simeq 0.672$ and energy is given in GeV. 

Furthermore, in order to take into account the dependence on the system size, we 
introduce a very simple $A$ dependence for the constant $T_0$ term. Looking at 
Fig.~\ref{tmunp} one can easily realize that $T_0$ depends almost logarithmically on the mass
number of the nucleus, thus it can be assumed $T_0 = T_c - \tau \log A$ with $T_c \simeq 191.5$ 
MeV and $\tau \simeq 4.5$ MeV leading to the final expression of the chemical 
freeze-out temperature:
\begin{equation}\label{tvsall}
  T =  0.1915 - 0.0045 \log A - 0.672 
  \left( \frac{\log \snn}{(\snn)^{1.13}} \right)^2 \,
\end{equation}
where every quantity is expressed in GeV. 

For $\gs$, as its dependence on $A$ is stronger than on $\snn$, it is more
suitable to write down an independent interpolation formula rather than obtaining
one from Eqs.~(\ref{tvsmu}), (\ref{gsvsmu}) and (\ref{muvse}). By using
\begin{equation}\label{gsvsall}
  \gs = 1 - \zeta \exp \left[ - \xi \sqrt{A \snn}\right]
\end{equation}
which is inspired of (\ref{muvse}), a good fit is obtained by setting $\zeta=0.606$
and $\xi = 0.0209$ (the energy is in GeV).  

It should be emphasized that all of the above relations only apply to
central collisions and cannot be used for peripheral nucleus-nucleus collisions.
For instance, it has already been observed, indeed, that strangeness under-saturation 
does not scale with the number of participants \cite{cley_systemsize},
rather with the linear size of the fireballs \cite{maiani}.

\section{Discussion} 

In principle, by using equations (\ref{tvsmu})-(\ref{gsvsall}), it is possible 
to estimate, within the SHM, the average multiplicities and ratios of any particle 
species for any colliding system in central collisions at any energy. It is 
especially interesting to study possible deviations of some specific particle 
ratios from the predicted smooth dependence of our interpolations. Such 
deviations would be a signal that the SHM scheme has some problem and either 
some refinement is needed (e.g. the hypotheses underlying the global fit do 
not fully apply) or there is some specific mechanism beyond the statistical 
{\rm ansatz} responsible for them. 

\begin{center}
\begin{figure}[!h]
$\begin{array}{c@{\hspace{1in}}c}
\epsfxsize=4.5in
\epsffile{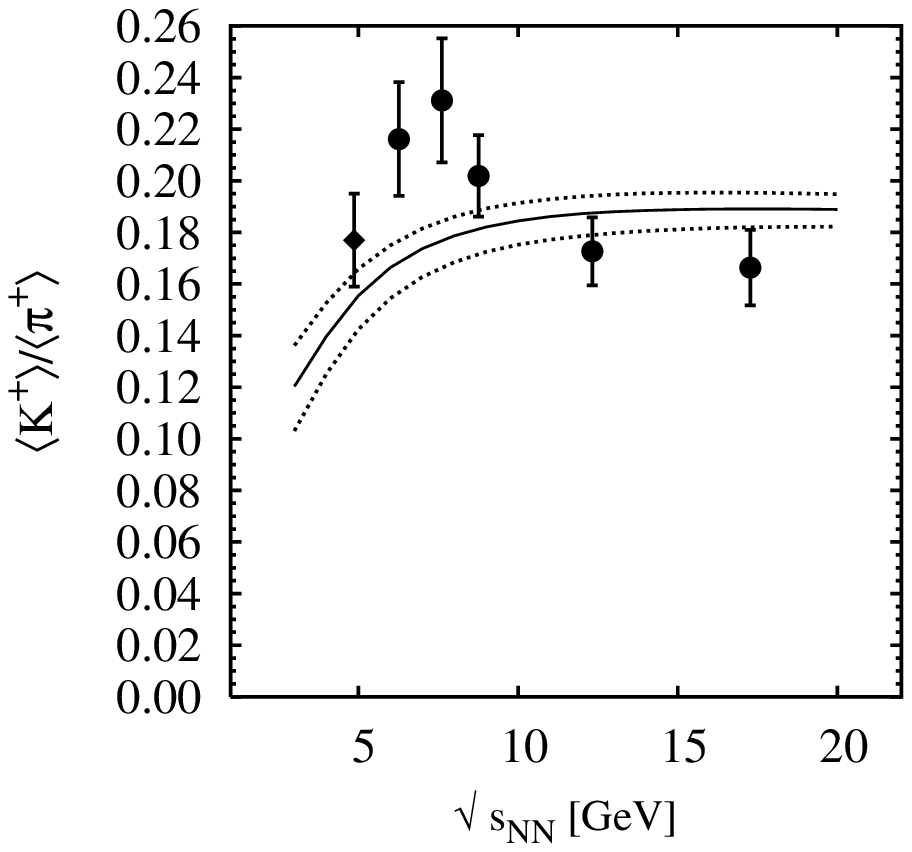} & \hspace*{-5cm}
        \epsfxsize=4.5in
        \epsffile{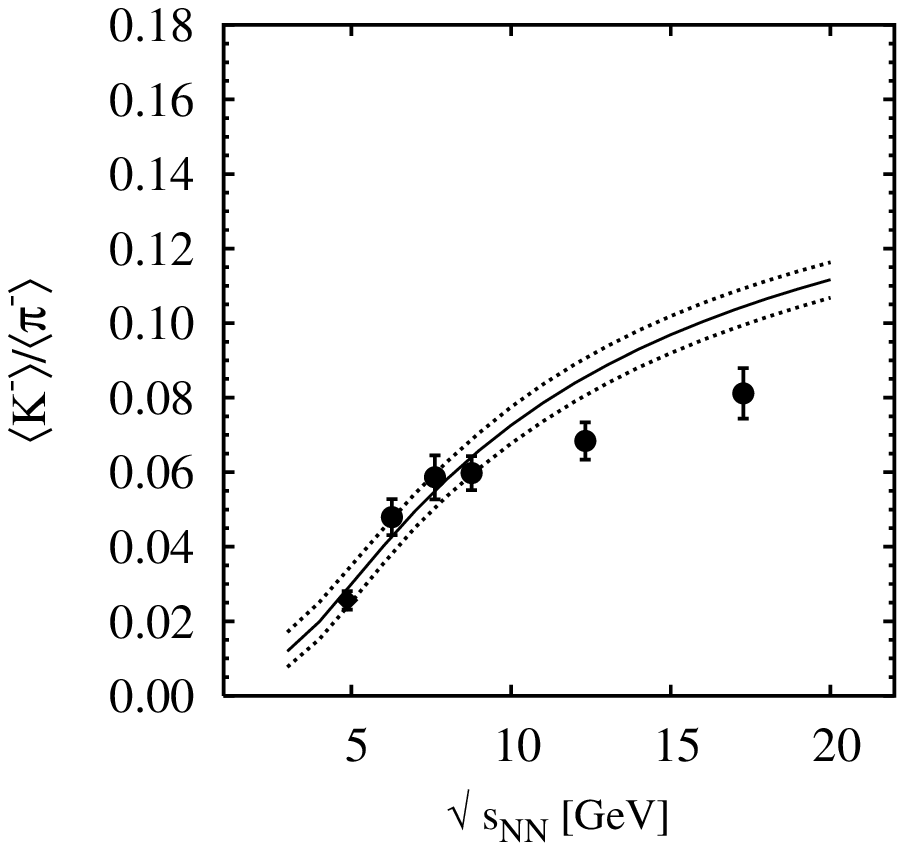} \\ 
\end{array}$
\vspace{-0.4cm}
\caption{Experimental vs interpolated \kpi~-ratios (LEFT) and \kmpim~-ratios
(RIGHT) in central heavy ion collisions as a function of $\snn$. The ``horn"
structure around $\snn \sim 7-8$ GeV is clearly visible. The bands have been 
calculated by using statistical model interpolating equations~(\ref{tvsmu}),
(\ref{muvse}),(\ref{gsvsall}) with their central best fit parameters and errors 
(see text).
\label{kpirats}}
\end{figure}
\end{center}

Of special importance in this context is the anomalous peak (``horn") in the 
ratio K$^+/\pi^+$ observed around beam energy 20$A$-30\agev~\cite{MarekPhi}, 
which has been discussed in the statistical model in ref.~\cite{cleyhorn}.
Fig.~\ref{kpirats} shows the experimental peak as a function of $\snn$ along 
with the statistical model prediction calculated by means of the interpolations 
(\ref{tvsmu})-(\ref{gsvsall}) and taking $A=208$ along the curve. The theoretical 
error band corresponds to the 1$\sigma$ ($\simeq 68 \%$) 7-dimensional ellipsoidal 
contour of the
multivariate Gaussian distribution relevant to the 7 free parameters in our 
interpolating relations (\ref{tvsmu}),(\ref{muvse}),(\ref{gsvsall}) (that is
$T_c$, $\tau$, $b$, $\alpha$, $\beta$, $\zeta$ and $\xi$). In practice, this band has 
been determined through a Monte-Carlo procedure by randomly extracting 1000 times 
these parameters within the above contour and calculating, for each set, the 
K$^+/\pi^+$ ratio; the resulting minimum and maximum ratios were taken as the 
band bounds. The smooth interpolation of the \SHMgs~parameters fails to reproduce 
the horn because the model (see Table~\ref{80158RHICmult}) underestimates $K^+$ 
multiplicity at each energy point over the relevant range, especially at 20$A$ and 
30\agev, where the discrepancy is of the order of 3 standard deviations. Pion 
multiplicities are also underestimated, but deviations are much less strong, 
especially at lower energies. In Fig.\ref{kpirats}, the corresponding \kmpim~ratio 
is shown as well. In Pb-Pb collisions, the statistical model tends to overestimate 
this ratio because K$^-$ multiplicities are relatively well reproduced at all 
beam energies, but the $\pi^-$ multiplicities are underestimated especially at 
the higher energies. 

There are many possible reasons for this and other observed discrepancies,
among them: 
\begin{itemize}
\item{} a failure of the \SHMgs~scheme;
\item{} an insufficient knowledge of the hadronic mass spectrum and branching 
 ratios;
\item{} an inadequacy of the assumptions needed to perform global fits,
e.g. large fluctuations of charge distributions among the different clusters.
\end{itemize}
As far as the first item is concerned, we observe that the most straightforward
way of overcoming these difficulties, is to add further parameters in the model, 
like the light-quark non-equilibrium parameter $\gq$ proposed in ref.~\cite{rafe}. 
However, unlike $\gs$, the introduction of this new parameter does not imply 
a remarkable and systematic improvement in the statistical model fits for all 
collisions \cite{bgs}. In fact, by making a careful scan of 4-parameters fit at 
20$A$ GeV, in the analysis A, we found that the best fit occurs at $\gq=0.7$, 
with $T \simeq 143$ MeV, $\mu_B/T \simeq 3.5$ with a $\chi^2/dof = 14.7/4$, 
i.e. only $0.8/4$ units$/dof$ less than the fit for 
$\gq=1$. On the other hand, at 30$A$ GeV, we obtain a better fit ($\chi^2/dof = 
9.5/4$ instead of $16.5/4$) at $\gamma_q = 1.7$ with $T \simeq 124$ MeV and 
$\mu_B/T \simeq 3$, in agreement with ref.~\cite{rafhorn}, at a price of an abrupt 
change of $\gq$, of about 1 and of a decrease (contrary to the general trend) of 
temperature of 20 MeV within a range of only 1.4 GeV in centre-of-mass energy. Such 
rapid changes might be simply owing to random fluctuations generated by data 
overfitting and definitely need further investigations. 

Concerning the second item, it should be emphasized that $\sim$70\% of $\pi^{\pm}$ 
and $\sim$50\% of K$^-$ multiplicities originate from resonance decays, while 
for K$^+$ the contribution increases from 25\% at AGS to $\sim$50\% at RHIC. 
Although the uncertainty on the measured branching ratios do not play a significant
role (see discussion in Sect.~3), the possible presence of many undetected 
resonances may affect the calculation of final yields. One of the most remarkable 
examples is the $\sigma$ scalar resonance which is usually not included in these 
analyses. In fact, if the mass was 600 MeV and width 300 MeV, its inclusion would 
entail an enhancement of charged pion yield by about 20 units in Pb-Pb collisions 
at $\snn=17.2$ GeV, which will make our fit better. However, it is difficult to
conclude that the inclusion of this resonance would restore a perfect agreement 
with the data with still such a large uncertainty on its parameters.

Altogether, we deem that, among the above proposed explanations of the observed 
discrepancies, the third one is the most conservative.
Non-statistical fluctuations of strangeness or baryon density, 
like those discussed e.g. in ref.~\cite{koch}, may invalidate the EGC assumption, 
thus affecting the global fit to particle multiplicities. Yet, this kind of 
alternative pictures still need to be probed with thorough comparison with the 
data. 

\begin{center}
\begin{figure}[ht]
\epsfxsize=6.5in
\epsffile{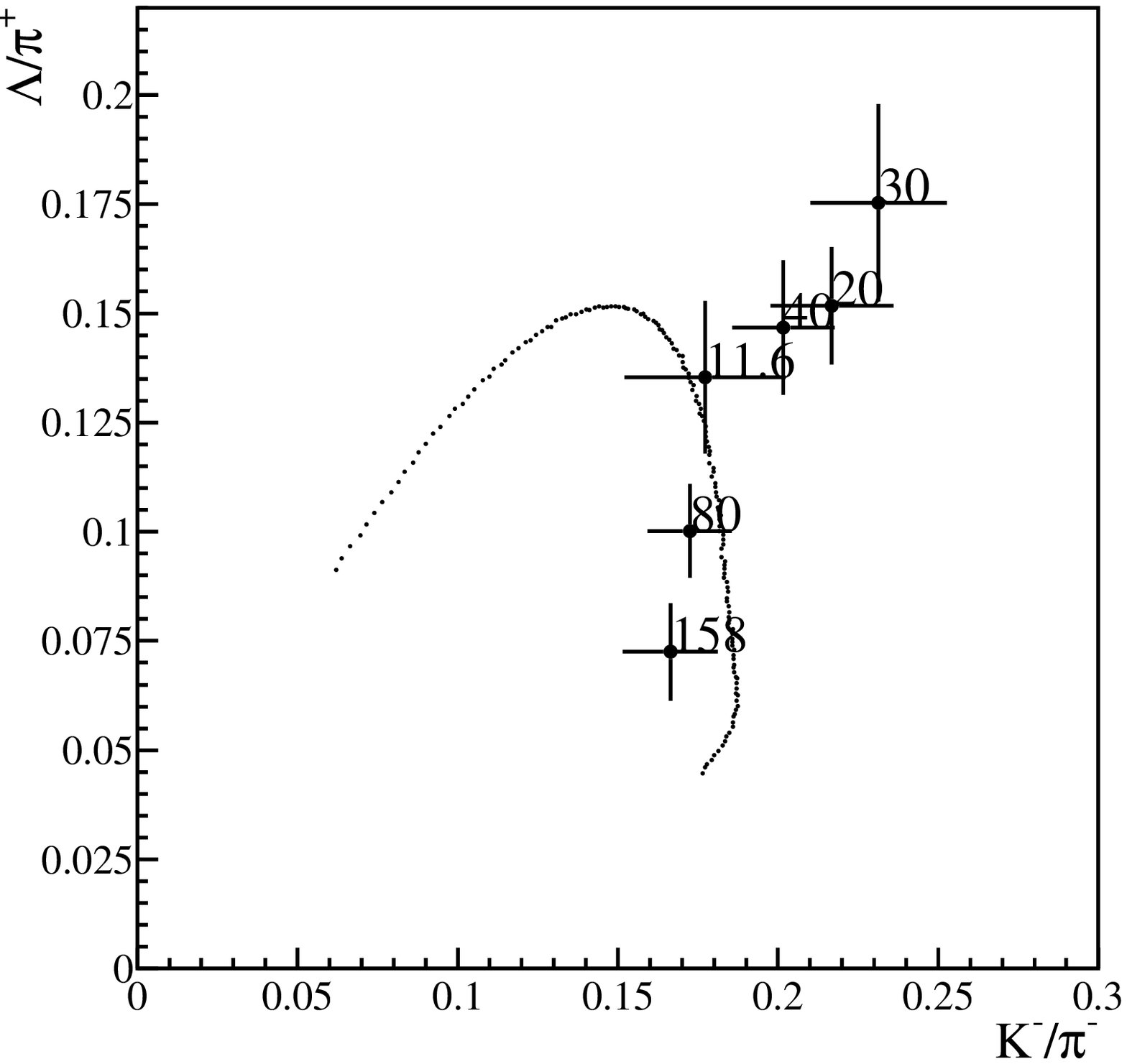} 
\caption{The $\Lambda/\pi^+$ vs K$^-/\pi^-$ ratios measured in  Pb-Pb and Au-Au collisions 
at various beam energies. The dashed line shows the central predicted values of
the interpolations (\ref{tvsmu}) and (\ref{gsvsmu}) with their best fit parameters.
\label{ratios1}}
\end{figure}
\end{center}

\begin{center}
\begin{figure}[ht]
\epsfxsize=6.5in
\epsffile{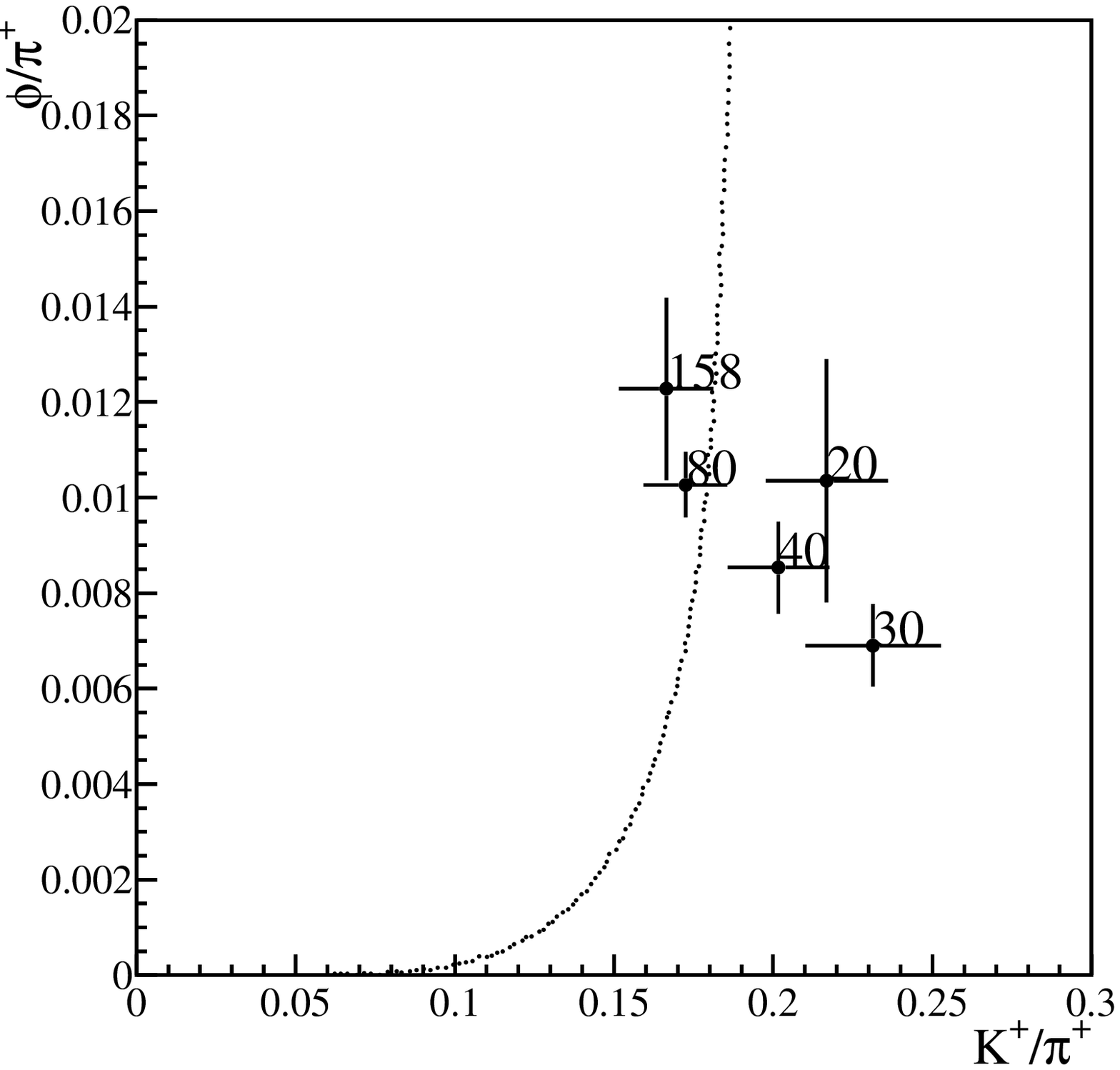} 
\caption{The $\phi/\pi^+$ vs K$^+/\pi^+$ ratios measured in  Pb-Pb and Au-Au collisions 
at various beam energies. The dashed line shows the central predicted values of
the interpolations (\ref{tvsmu}) and (\ref{gsvsmu}) with their best fit parameters.
\label{ratios2}}
\end{figure}
\end{center}

\begin{center}
\begin{figure}[ht]
\epsfxsize=6.5in
\epsffile{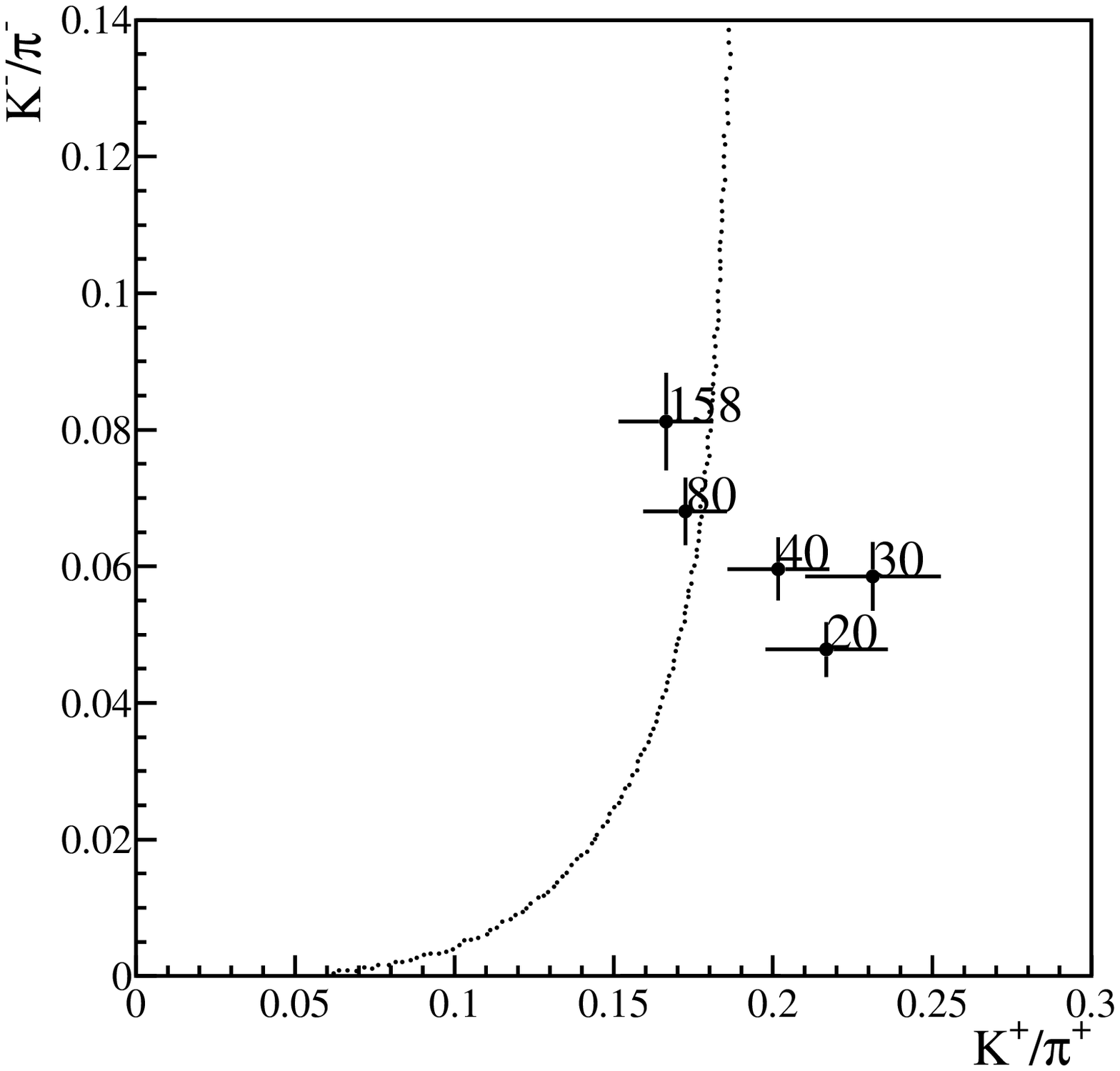} 
\caption{The K$^+/\pi^+$ vs K$^-/\pi^-$ ratios measured in Pb-Pb and Au-Au collisions 
at various beam energies. The dashed line shows the central predicted values of
the interpolations (\ref{tvsmu}) and (\ref{gsvsmu}) with their best fit parameters.
\label{ratios3}}
\end{figure}
\end{center}

The collisions between 20$A$ and 40\agevt of beam energy are those showing the most
significant discrepancies from the smooth SHM interpolations. This can be seen in 
Figs.~\ref{ratios1}, \ref{ratios2} and \ref{ratios3}, where the correlations between 
different particle ratios are shown along with the predicted central values of the 
interpolations (\ref{tvsmu}), (\ref{gsvsmu}) and (\ref{gsvsall}) as smooth dashed
lines. All of these ratios involve strangeness-carrying particles, which may suggest 
that a peculiar dynamical process involving strange quarks occurs around this 
energy region. 

\section{Summary and Conclusions}\label{summary}

We have analyzed the available hadronic multiplicities measured in central 
heavy ion collisions over an energy range $\snn = 4.5 - 17.2$ GeV within 
the statistical hadronization model. The thermal parameters of the source,
baryon chemical potential and temperature, depend strongly on N-N 
centre-of-mass energy, but their behaviour is found to be  smooth
and we have been able to find  empirical relations describing their 
dependence on energy up to the top RHIC energy. Conversely, at fixed energy, they 
depend  little on the system size as we have found similar values for 
$\mu_B$ and $T$ at chemical freeze-out for C-C, Si-Si and Pb-Pb. 

On the other hand, chemical equilibration of strangeness is seen to be more
dependent on the number of colliding nucleons than on energy, with the general
trend for the strangeness under-saturation parameter $\gs$ to attain the value
1 only at energies $\snn = {\cal O}(100)$ GeV. Different versions of the SHM,
aiming at explaining the observed under-saturation of the strange particle 
phase space, namely the two-component model described here and strangeness 
correlation volume model examined in ref.~\cite{last}, seem to be disfavored 
by the data.
 
Discrepancies between the model with strangeness under-saturation parameter 
and the data have been pointed out with regard to specific particle ratios.
The origin of these discrepancies, mostly in the beam energy region 20$A$-40\agev,
is still to be investigated.
  
\section*{Acknowledgments}

This work was in partly supported by the NordForsk (J.~M.) and Virtual Institute 
of Strongly Interacting Matter (VI-146) of Helmholtz Association, Germany. 



\newpage 
\begin{table}
\begin{center}
\caption{Comparison between measured and fitted particle multiplicities, in the
framework of SHM($\gs$) model, in p-p and central C-C (15.3\% most central) and 
Si-Si (12.2\%) collisions at a beam energy of 158$A$ GeV.
}\label{Smallmult}
\vspace{0.5cm}
\begin{tabular}{|c|c|c|c|c|c|c|c|c|c|}

\hline
\multicolumn{1}{|c|}{} & \multicolumn{3}{|c|}{p-p 158\agevt  (canonical)} & 
\multicolumn{3}{|c|}{C-C 158\agevt (S-canonical)} & \multicolumn{3}{|c|}{Si-Si 158\agevt (S-canonical)} \\
\hline
                   & Measurement            & Fit A  &  Fit B & Measurement & Fit A  &  Fit B & Measurement & Fit A  &  Fit B \\
\hline                                                                                           
 $N_P$             &                                           &         &   
                   & $16.3 \pm 1 $~\cite{NA49CCSiSi}           & 15.79   &  15.98 
                   & $41.4 \pm 2 $~\cite{NA49CCSiSi}           & 39.87   &  40.30 \\
 $\pi^+$           & 3.15$\pm$0.16~\cite{NA49pp_all}           & 3.25     & 3.28
                   & $22.4 \pm 1.6$~\cite{NA49CCSiSi}  & 22.3     & 22.0  
                   & $56.6 \pm 4.1$~\cite{NA49CCSiSi}  & 57.4     & 56.47  \\
 $\pi^-$           & 2.45$\pm$0.12~\cite{NA49pp_all}           & 2.43     &  2.45
                   & $22.2 \pm 1.6$~\cite{NA49CCSiSi}  & 22.3     &  22.0 
                   & $57.6 \pm 4.1$~\cite{NA49CCSiSi}  & 57.5     &  56.53 \\
 K$^+$             & 0.21$ \pm$0.02~\cite{NA49pp_all}          & 0.228     &  0.200   
                   & $2.54 \pm 0.25$~\cite{NA49CCSiSi} & 2.71      &  2.79
                   & $7.44 \pm 0.74$~\cite{NA49CCSiSi} & 7.99      &  8.17 \\     
 K$^-$             & 0.13$\pm$0.013~\cite{NA49pp_all}         & 0.119     & 0.107 
                   & $1.49 \pm 0.16$~\cite{NA49CCSiSi} & 1.46      & 1.51
                   & $4.42 \pm0.44$~\cite{NA49CCSiSi} & 4.32      & 4.49  \\     
 $\phi$            & 0.012$\pm$0.0015~\cite{NA49pp_all}     & 0.0203$^{a}$ &  0.0149     
                   & $0.18 \pm0.02$~\cite{NA49CCSiSi} & 0.15         &  0.16    
                   & $0.66 \pm0.09$~\cite{NA49CCSiSi} & 0.48         &  0.51 \\ 
$\Lambda$          & 0.115$\pm$0.012~\cite{NA49pp_all}          & 0.133      &  0.117     
                   & $1.32 \pm 0.32$~\cite{NA49CCSiSi} & 1.62       &  1.69
                   & $3.88 \pm 0.58$~\cite{NA49CCSiSi} & 4.57       &  4.87 \\ 
$\bar\Lambda$      & 0.0148$\pm$0.0019~\cite{NA49pp_all}           & 0.0147  & 0.0141    
                   & $0.177 \pm 0.028$~\cite{Kraus:2004uk} & 0.149     & 0.182    
                   & $0.492 \pm 0.108$~\cite{Kraus:2004uk} & 0.389    & 0.508 \\
 $\Xi^-$           & 0.0031$\pm$0.0003~\cite{NA49pp_all}           & 0.00285 & 0.00110$^{a}$
                   &   & 0.0728 & 0.0666           
                   &   & 0.257 & 0.244       \\  
 $\bar\Xi^+$       & $9.2\,10^{-4}\pm0.9\,10^{-5}$~\cite{NA49pp_all} & 0.000918  & 0.000388$^{a}$
                   & & 0.0151   & 0.0161
                   & & 0.0485  & 0.0562  \\
 $\Omega$          & $2.6\,10^{-4}\pm1.3\,10^{-4}$~\cite{NA49pp_all} & $8.87 \,10^{-5}$ & $2.12 \,10^{-5}$$^{a}$
                   & & 0.00397     & 0.00405
                   & & 0.0181     & 0.0196  \\
 $\bar\Omega$      & $1.6\,10^{-4}\pm0.9\,10^{-4}$~\cite{NA49pp_all} & $6.16 \,10^{-5}$ & $1.30 \,10^{-5}$$^{a}$
                   & & 0.00179     & 0.00216
                   & & 0.00736    & 0.00979  \\
 p                 & & 1.094    & 1.125
                   & & 7.01     & 7.18
                   & & 17.2     & 17.6  \\
 $\bar{\rm p}$     & 0.040$\pm$0.007~\cite{NA49pp_all} & 0.0364  & 0.0445
                   &                                   & 0.303     & 0.367 
                   &                                   & 0.714    & 0.879  \\
 K$^0_S$      	   & 0.18$\pm$0.04~\cite{NA49pp_all} & 0.14     & 0.15
                   &                                 & 2.05       & 2.14
                   &                                 & 6.06       & 6.37  \\
 $\pi^0$      	   & & 3.32     & 3.10
                   & & 25.0     & 23.2
                   & & 64.3     & 59.7  \\
 $\eta$       	   & & 0.382    & 0.279
                   & & 2.64     & 2.09
                   & & 6.78     & 5.59  \\
 $\omega$     	   & & 0.342    & 0.299
                   & & 2.33     & 2.00
                   & & 5.85     & 5.04  \\
 $\eta^{'}$    	   & & 0.0328   & 0.0165
                   & & 0.210     & 0.132
                   & & 0.537    & 0.374  \\
 $\rho^+$     	   & & 0.449    & 0.467
                   & & 2.75     & 2.84
                   & & 6.92     & 7.11  \\
 $\rho^-$     	   & & 0.301    & 0.305
                   & & 2.75     & 2.85
                   & & 6.94     & 7.16  \\
 $\rho^0$     	   & & 0.408    & 0.428
                   & & 2.84     & 2.94
                   & & 7.14     & 7.39  \\
 K$^{*+}$     	   & & 0.0878   & 0.0673
                   & & 0.978     & 0.933
                   & & 2.80     & 2.72  \\
 K$^{*-}$     	   & & 0.0359   & 0.0275
                   & & 0.462     & 0.437
                   & & 1.35     & 1.29  \\
 K$^{*0}$     	   & & 0.0741   & 0.0563
                   & & 0.964     & 0.926
                   & & 2.76     & 2.70  \\
$\bar{\rm K}^{*0}$ & & 0.0405   & 0.0316
                   & & 0.455    & 0.431
                   & & 1.33     & 1.27  \\
 $\Delta^{++}$     & & 0.281    & 0.263
                   & & 1.62     & 1.52
                   & & 3.92     & 3.69  \\
 $\bar\Delta^{--}$ & & 0.0063    & 0.00701
                   & & 0.0686    & 0.0748
                   & & 0.160     & 0.177  \\
 $\Sigma^+$        & & 0.0413   & 0.0321
                   & & 0.444    & 0.429
                   & & 1.26     & 1.24  \\
 $\Sigma^-$        & & 0.0276   & 0.0213
                   & & 0.435    & 0.421
                   & & 1.23     & 1.21  \\
 $\Sigma^0$        & & 0.0358   & 0.0276
                   & & 0.440    & 0.424
                   & & 1.25     & 1.22  \\
 $\bar\Sigma^-$    & & 0.00339   & 0.00295
                   & & 0.0413    & 0.0472
                   & & 0.111     & 0.131  \\
 $\bar\Sigma^+$    & & 0.00445   & 0.00388
                   & & 0.0403    & 0.0461
                   & & 0.107     & 0.128  \\
 $\bar\Sigma^0$    & & 0.00407   & 0.00352
                   & & 0.0409    & 0.0466
                   & & 0.109     & 0.130  \\
 $\Xi^0$           & & 0.00323   & 0.00126
                   & & 0.0740    & 0.0676
                   & & 0.262     & 0.248  \\
 $\bar\Xi^0$       & & 0.000849  & 0.000351
                   & & 0.0152    & 0.0164
                   & & 0.0494    & 0.0572  \\
 $\Lambda(1520)$   & 0.012$\pm$0.003~\cite{NA49pp_all} & 0.0106       & 0.00787
                   &                                   & 0.116        & 0.110
                   &                                   & 0.321        & 0.313  \\
\hline 
\multicolumn{10}{|l|}{$a$ - Excluded from the data sample in the fit}.\\
\hline
\end{tabular}
\end{center}
\end{table}

\newpage 
\begin{table}
\begin{center}
\caption{Comparison between measured and fitted particle multiplicities, in the framework of
SHM($\gs$) model, in central Au-Au collisions (3\%) at 11.6$A$ GeV as well as Pb-Pb collisions (7\%) at 
beam energies of 20$A$ and 30$A$ 
GeV.}\label{203040mult}
\vspace{0.5cm}
\begin{tabular}{|c|c|c|c|c|c|c|c|c|c|}
\hline
\multicolumn{1}{|c|}{} & \multicolumn{3}{|c|}{Au-Au 11.6\agevt  (GC ensemble)} & 
\multicolumn{3}{|c|}{Pb-Pb 20\agevt (GC ensemble)} & \multicolumn{3}{|c|}{Pb-Pb 30\agevt (GC ensemble)} \\
\hline
                   & Measurement            & Fit A  &  Fit B & Measurement 
& Fit A  &  Fit B & Measurement & Fit A  &  Fit B \\
\hline                                                                                           
 $N_P$             & $363 \pm  10$~\cite{centrags}          & 365.6     &    360.6
                   & $349 \pm  1 \pm 5$~\cite{MarekPhi}     & 347.5     &    347.3 
                   & $349 \pm  1 \pm 5$~\cite{AltPbPb2004}  & 350.2     &    350.2 \\
 $\pi^+$           & $133.7 \pm 9.93$~\cite{pionags,beca01} & 135.7      &  129.4
                   & $184.5\pm 0.6\pm 13$~\cite{MarekPhi}   & 193.7      &  193.3 
                   & $239\pm 0.7\pm 17$~\cite{AltPbPb2004}  & 247.5      &  254.3  \\
 $\pi^-$           &                                       & 177.7     &  156.1
                   & $217.5\pm 0.6\pm 15$~\cite{MarekPhi}  & 221.4     &  220.7 
                   & $275\pm 0.7\pm 19$~\cite{AltPbPb2004} & 276.2     &  283.2  \\
 K$^+$             & $23.7 \pm 2.86$~\cite{centrags}        & 18.7     &  18.8
                   & $40.0\pm 0.8\pm 2.0$~\cite{MarekPhi}   & 34.4     &  35.7 
                   & $55.3\pm 1.6\pm 2.8$~\cite{AltPbPb2004}& 44.8     &  43.9 \\
 K$^-$             & $3.76 \pm 0.47$~\cite{centrags}        & 3.90     &  3.54 
                   & $10.4\pm 0.12\pm 0.5$~\cite{MarekPhi}  & 10.5     &  10.4 
                   & $16.1\pm 0.2\pm 0.8$~\cite{AltPbPb2004}& 16.3     &  16.1 \\
 $\phi$            &                                   & 0.327    & 0.350    
                   & $1.91\pm 0.45$~\cite{MarekPhi}    & 1.20     & 1.44     
                   & $1.65\pm 0.17$~\cite{AltPbPb2004} & 2.01     & 2.08    \\ 
$\Lambda$          & $18.1 \pm 1.9$~\cite{lamb891,lamb896,last} & 19.7     & 19.6
                   & $28.0\pm 1.5$~\cite{PbPbhyper}                 & 28.8     &  31.2   
                   & $41.9\pm 2.1\pm 4.0$~\cite{PbPbhyper}          & 33.8     &  34.5 \\
$\bar\Lambda$      & $0.017 \pm 0.005$~\cite{lamb891,lamb896,last} & 0.017    & 0.011
                   & $0.16\pm 0.03$~\cite{BlumeSQM2004}                & 0.11     & 0.16     
                   & $0.50\pm 0.04$~\cite{PbPbhyper}                   & 0.35     & 0.49  \\  
$p/\pi^+$          & $1.23 \pm 0.13$~\cite{protags,beca01} & 1.27      & 1.22
                   &                                       &           &
                   &                                       &           & \\
 $\Xi^-$           &                                          & 0.551 & 0.557   
                   &                                          & 1.37  & 1.42
                   &                                          & 1.85  & 1.66    \\
 $\bar\Xi^+$       &                                      & 0.00221   & 0.00244   
                   &                                      & 0.0215    & 0.0340
                   &                                      & 0.0645    & 0.0846 \\
 $\Omega$          & & 0.0132   & 0.0147
                   & & 0.0618   & 0.0758
                   & & 0.105    & 0.104 \\
 $\bar\Omega$      & & 0.000367  & 0.000571
                   & & 0.00436   & 0.00936
                   & & 0.0134    & 0.0204 \\
 p                 & & 172.5     & 155.1
                   & & 143.6     & 142.0
                   & & 142.0     & 142.6 \\
 $\bar{\rm p}$     & & 0.0302   & 0.0124
                   & & 0.144    & 0.170
                   & & 0.467    & 0.625 \\
 K$^0_S$      	   & & 11.6     & 11.7
                   & & 22.7     & 23.8
                   & & 30.7     & 30.6 \\
 $\pi^0$      	   & & 164.5     & 146.7
                   & & 224.2     & 215.3
                   & & 286.2     & 280.8 \\
 $\eta$       	   & & 8.17     & 6.83
                   & & 16.7     & 15.2
                   & & 24.5     & 21.8 \\
 $\omega$     	   & & 5.03     & 3.62
                   & & 11.3     & 9.50
                   & & 17.8     & 15.9 \\
 $\eta^{'}$    	   & & 0.335    & 0.243
                   & & 0.938    & 0.818
                   & & 1.58     & 1.30 \\
 $\rho^+$     	   & & 7.74     & 10.3
                   & & 15.2     & 18.5
                   & & 22.5     & 26.7 \\
 $\rho^-$     	   & & 9.19     & 12.6
                   & & 17.5     & 21.5
                   & & 25.5     & 30.4 \\
 $\rho^0$     	   & & 8.56     & 11.5
                   & & 16.6     & 20.4
                   & & 24.6     & 29.2 \\
 K$^{*+}$     	   & & 3.59     & 3.50
                   & & 8.38     & 8.93
                   & & 12.4     & 12.2 \\
 K$^{*-}$     	   & & 0.627    & 0.513
                   & & 2.19     & 2.07
                   & & 3.93     & 3.67 \\
 K$^{*0}$     	   & & 3.80     & 3.81
                   & & 8.75     & 9.50
                   & & 12.9     & 12.8 \\
$\bar{\rm K}^{*0}$ & & 0.564   & 0.459
                   & & 2.01    & 1.89
                   & & 3.65    & 3.41 \\
 $\Delta^{++}$     & & 25.6     & 24.2
                   & & 26.5     & 25.4
                   & & 27.9     & 26.8 \\
 $\bar\Delta^{--}$ & & 0.00330   & 0.00222
                   & & 0.0296    & 0.0332
                   & & 0.100     & 0.126  \\
 $\Sigma^+$        & &  4.81   & 4.75
                   & &  7.54   & 7.65
                   & &  8.92   & 8.46 \\
 $\Sigma^-$        & &  5.43   & 5.48
                   & &  8.30   & 8.49
                   & &  9.67   & 9.21 \\
 $\Sigma^0$        & &  5.13   & 5.10
                   & &  7.94   & 8.05
                   & &  9.32   & 8.82 \\
 $\bar\Sigma^-$    & &  0.00363 & 0.00323
                   & &  0.0329  & 0.0449
                   & &  0.103   & 0.136  \\
 $\bar\Sigma^+$    & &  0.00295  & 0.00257
                   & &  0.0278   & 0.0379
                   & &  0.0889   & 0.118  \\
 $\bar\Sigma^0$    & &  0.00328  & 0.00288
                   & &  0.0303   & 0.0412
                   & &  0.0959   & 0.127 \\
 $\Xi^0$           & &  0.537   & 0.535
                   & &  1.34    & 1.39
                   & &  1.83    & 1.62 \\
 $\bar\Xi^0$       & &  0.00248  & 0.00270
                   & &  0.0231   & 0.0366
                   & &  0.0686   & 0.0899 \\
 $\Lambda(1520)$   & &  0.776   & 0.777
                   & &  1.47    & 1.58
                   & &  1.94    & 1.91 \\
\hline 
\end{tabular}
\end{center}
\end{table}

\newpage 
\begin{table}
\begin{center}
\caption{Comparison between measured and fitted particle multiplicities, in the framework of
SHM($\gs$) model, in central Pb-Pb collisions at beam energies of 40 and 80 (7\% most central) and 
158$A$ GeV (5\% most central). The measured $\Lambda(1520)$ yield has been removed from the 
fitted data sample at 158\agev.}\label{80158RHICmult}
\vspace{0.5cm}
\begin{tabular}{|c|c|c|c|c|c|c|c|c|c|}
\hline

\multicolumn{1}{|c|}{} & \multicolumn{3}{|c|}{Pb-Pb 40\agevt (GC ensemble)} 
&\multicolumn{3}{|c|}{Pb-Pb 80\agevt  (GC ensemble)} & 
\multicolumn{3}{|c|}{Pb-Pb 158\agevt (GC ensemble)}  \\
\hline
                   & Measurement            & Fit A  &  Fit B & Measurement & Fit A  &  
Fit B & Measurement & Fit A  &  Fit B \\
\hline                                                                                           
 $N_P$             & $349 \pm  1 \pm  5$~\cite{NA49Afan}    & 351.4     &    351.2 
                   & $349 \pm  1 \pm 5$~\cite{NA49Afan}  & 351.2   &  351.0
                   & $362 \pm 1 \pm  5$~\cite{NA49Afan}  & 363.2   &  363.4 \\
$\pi^+$            & $293 \pm 3 \pm 15$~\cite{NA49Afan}    & 283.4      &  285.1  
                   & $446 \pm 5 \pm 22$~\cite{NA49Afan}         & 420.1   & 419.4
                   & $619 \pm 17 \pm 31$~\cite{NA49Afan}        & 550.0   & 533.6 \\ 
 $\pi^-$           & $322 \pm 3 \pm 16$~\cite{NA49Afan}    & 312.6     &  314.6 
                   & $474 \pm 5 \pm 23$~\cite{NA49Afan}        & 450.6   & 450.4 
                   & $639 \pm 17 \pm 31$~\cite{NA49Afan}        & 582.0   & 565.9   \\
K$^+$              & $59.1 \pm 1.9 \pm 3$~\cite{NA49Afan}   & 52.1     &  52.1      
                   & $76.9 \pm 2 \pm 4$~\cite{NA49Afan}     & 71.6     & 71.0 
                   & $103 \pm 5 \pm 5$~\cite{NA49Afan}       & 103.1    & 103.6 \\
 K$^-$             & $19.2 \pm 0.5 \pm 1.0$~\cite{NA49Afan} & 20.8     &  20.8      
                   &  $32.4 \pm 0.6 \pm 1.6$~\cite{NA49Afan} & 36.5     & 36.4 
                   & $51.9 \pm 1.9 \pm 3$~\cite{NA49Afan}    & 59.3     & 59.3 \\
 $\phi$            & $2.50 \pm 0.25$~\cite{MarekPhi}   & 2.70     & 2.77 
                   &  $4.58 \pm 0.20$~\cite{MarekPhi}     & 4.43    & 4.46     
                   & $7.6 \pm 1.1$~\cite{NA49Afan}        & 8.01    & 8.59  \\
$\Lambda$          & $43.0 \pm 1.9\pm3.4$~\cite{Anticic2004} & 37.1     &  38.2  
                   &  $44.7 \pm 2.8\pm3.2$~\cite{Anticic2004} & 42.4     & 43.2       
                   & $44.9\pm3.5\pm 5.4$~\cite{Anticic2004}   & 53.6     & 55.9 \\
$\bar\Lambda$      & $0.66 \pm 0.04\pm0.06$~\cite{Anticic2004} & 0.67     & 0.64 
                   &  $2.02\pm 0.25\pm0.20$~\cite{Anticic2004} & 2.16   & 2.11     
                   & $3.74\pm 0.19\pm0.43$~\cite{Anticic2004}  & 4.83   & 4.96  \\  
 $\Xi^-$           & $2.41 \pm0.15\pm0.24$~\cite{PbPbhyper2}  & 2.20  & 2.10       
                   &                                           &  2.83    & 2.62
                   & $4.45 \pm 0.22$~\cite{NA49Afan}           &  4.42    & 4.31  \\
 $\bar\Xi^+$       &                                      & 0.122    & 0.112 
                   &                                      & 0.330    & 0.298
                   & $0.83 \pm 0.04$~\cite{NA49Afan}      & 0.79     & 0.78 \\
 $\Omega$          &                                      & 0.143    & 0.148  
                   &                                      & 0.222    & 0.221
                   & $0.59\pm 0.12\pm0.04$~\cite{Alt2004} & 0.44     & 0.49 \\
 $\bar\Omega$      &                                      & 0.0262   & 0.0268  
                   &                                      & 0.0623   & 0.0610
                   & $0.26\pm 0.06\pm0.03$~\cite{Alt2004} & 0.16     & 0.19 \\
$\Omega+\bar\Omega$  & $0.14 \pm 0.03 \pm 0.04$~\cite{Alt2004}& 0.17      & 0.18 
                     &                                        &           & 
                     &                                        &           & \\  
                     
 p                 & & 140.8     & 140.7  
                   & & 140.8     & 141.3
                   & & 143.4     & 142.8 \\
  $\bar{\rm p}$    &                                   & 0.897    & 0.821 
                   &                                   & 3.32     & 3.21
                   &                                   & 6.86     & 6.70  \\
 K$^0_S$           &                               & 36.4     & 37.1 
      	           &                               & 53.5     & 54.4
                   & $81 \pm 4$~\cite{Mischke2003} & 80.0     & 82.6 \\
 $\pi^0$           & & 328.1     & 314.1  
     	           & & 485.0     & 458.1
                   & & 636.2     & 581.6 \\
 $\eta$            & & 30.2     & 26.1  
         	   & & 49.5     & 41.6
                   & & 70.5     & 62.2 \\
  $\omega$         & & 22.7     & 18.5  
     	           & & 39.9     & 33.0
                   & & 55.4     & 44.7 \\
 $\eta^{'}$        & & 2.11     & 1.60  
    	           & & 3.76     & 2.75
                   & & 5.73     & 4.41 \\
 $\rho^+$          & & 27.9     & 29.7  
    	           & & 46.7     & 47.2
                   & & 63.4     & 60.9 \\
 $\rho^-$          & & 31.3     & 33.5  
      	           & & 50.9     & 51.8
                   & & 68.2     & 66.1 \\
$\rho^0$           & & 30.3     & 32.5  
      	           & & 50.2     & 51.1
                   & & 68.0     & 66.1 \\
 K$^{*+}$          & & 15.6     & 14.3  
     	           & & 23.0     & 20.9
                   & & 34.2     & 31.8 \\
 K$^{*-}$          & & 5.41     & 4.80  
     	           & & 10.6     & 9.46
                   & & 18.0     & 16.2 \\
 K$^{*0}$          & & 15.6     & 14.9  
     	           & & 23.5     & 21.6
                   & & 34.7     & 32.5 \\
$\bar{\rm K}^{*0}$ & & 5.06    & 4.47  
                   & & 10.0     & 8.95
                   & & 17.2     & 15.5 \\
  $\Delta^{++}$    & & 28.7     & 26.4  
                   & & 29.9     & 27.5
                   & & 30.9     & 28.1 \\
 $\bar\Delta^{--}$ & & 0.198     & 0.163  
                   & & 0.747    & 0.650
                   & & 1.55     & 1.36 \\
 $\Sigma^+$        & &  9.79   & 9.47  
                   & & 11.3     & 10.8
                   & & 14.4     & 14.0 \\
 $\Sigma^-$        & &  10.5   & 10.2 
                   & & 11.9     & 11.4
                   & & 15.0     & 14.6 \\
 $\Sigma^0$        & &  10.2   & 9.83  
                   & & 11.6     & 11.1
                   & & 14.7     & 14.3 \\
 $\bar\Sigma^-$    & &  0.197   & 0.179  
                   & & 0.623    & 0.577
                   & & 1.38     & 1.34 \\
 $\bar\Sigma^+$    & &  0.172    & 0.157  
                   & & 0.561    & 0.520
                   & & 1.26     & 1.23 \\
 $\bar\Sigma^0$    & &  0.185    & 0.167  
                   & & 0.592    & 0.547
                   & & 1.32     & 1.28 \\
  $\Xi^0$          & &  2.17    & 2.07  
                   & & 2.81     & 2.60
                   & & 4.41     & 4.29 \\
  $\bar\Xi^0$      & &  0.129    & 0.119 
                   & & 0.345    & 0.312
                   & & 0.823    & 0.812 \\
 $\Lambda(1520)$   &                                   & 2.27    & 2.09  
                   &                                   & 2.78     & 2.53
                   & $1.57 \pm 0.44$~\cite{Friese2002} & 3.64     & 3.41 \\
\hline 
\end{tabular}
\end{center}
\end{table}


\clearpage

\vspace{-5cm}
\vspace{-5cm}
\vspace{-5cm}
\vspace{-5cm}

\end{document}